\documentclass{lmcs} 
\pdfoutput=1

\usepackage{lastpage}
\lmcsdoi{18}{1}{16}
\lmcsheading{}{\pageref{LastPage}}{}{}%
{Dec.~21,~2020}{Jan.~19,~2022}{}

\pdfoutput=1


\def\eg{{\em e.g.}}
\def\cf{{\em cf.}}
\def\ie{{\em i.e.}}

\usepackage[utf8]{inputenc}
\usepackage{graphicx}
\usepackage{amssymb}
\usepackage{stmaryrd}  
\usepackage{mathtools} 
\usepackage[english]{babel}

\usepackage{url}
\allowdisplaybreaks
\usepackage[misc,geometry]{ifsym}
\usepackage{tikz}
\usepackage{tikz-cd}
\usetikzlibrary{calc,arrows}
\usetikzlibrary{decorations.pathmorphing}
\usepackage{proof}

\newcommand{\Lb}{L^\bot} 
\newcommand{\Lh}{{\mathbf{\Gamma}}} 
\newcommand{\Lo}{\mathbf{L}} 
\newcommand{\mm}{^{-}} 
\renewcommand{\cc}{^\circ} 
\newcommand{\pp}{^{+}} 

\newcommand{\domM}{\ensuremath{\mathrm{\dom(-)}}}  
\newcommand{\domP}{\ensuremath{\mathrm{\dom(+)}}}  
\newcommand{\dom}{\ensuremath{\mathrm{dom}}} 
\newcommand{\supp}{\ensuremath{\mathrm{supp}}} 

\newcommand{\upset}{\ensuremath{\mathord{\uparrow}\mkern1mu}} 
\newcommand{\downset}{\ensuremath{\mathord{\downarrow}\mkern1mu}} 
\newcommand{\mee}{\wedge}
\newcommand{\bigmee}{\bigwedge}

\newcommand\ARG{\mkern1.5mu\text{-}\mkern1.5mu}
\newcommand{\ev}{\mathrm{ev}}

\newcommand{\inv}{^{-1}} %
\newcommand\sue{\subseteq} %
\newcommand{\N}{\ensuremath{\mathbb{N}}}
\newcommand{\Z}{\ensuremath{\mathbb{Z}}}
\newcommand{\Q}{\ensuremath{\mathbb{Q}}}
\newcommand{\ui}{[0,1]}
\newcommand{\w}{\widehat}

\newcommand{\mip}{-} 
\newcommand{\miss}{\ensuremath{\sim}} 

\newcommand\ic{\iota}
\newcommand\im{\iota_{-}}

\newcommand\MBU[2]{\ensuremath{\sem[#1 \geq #2]}} 
\newcommand\MBD[2]{\ensuremath{\sem[#1 < #2]}}    

\newcommand\NOdM{Ne\v{s}et\v{r}il and Ossona de Mendez}

\newcommand{\defiff}{\enspace \stackrel{\mathrm{def}}{\Longleftrightarrow}\enspace}

\newcommand\can{^\delta}
\newcommand\M{\mathcal M^\infty}
\newcommand\J{\mathcal J^\infty}
\newcommand\I{\mathcal I}
\newcommand\F{\mathcal F}

\newcommand\ki{\kappa\inv}
\newcommand\la{^\flat}
\newcommand\ra{^\#}
\newcommand{\zer}{@}

\newcommand\leftplus{\mathbin{\oplus^{\pi\mkern2mu\flat}}}

\newcommand{\opi}{\oplus^\pi} 

\newcommand\ee[1]{\enspace #1 \enspace}
\newcommand\ete[1]{{\enspace\text{#1}\enspace}}
\newcommand\qtq[1]{{\quad\text{#1}\quad}}

\newcommand\FO{\mathrm{FO}}
\newcommand\Fin{\mathrm{Fin}}
\newcommand\Fs{{\mathcal{F}}}
\newcommand\SP[1]{\left< #1 \right>_{\mathrm{\Gamma}}}
\newcommand\SPc[1]{\left< #1 \right>_{\mathrm{I}}}
\newcommand\SPp[1]{\left< #1 \right>} 
\newcommand\PrG[1]{\mathbb{P}_{\geq #1}\,}
\newcommand\PrL[1]{\mathbb{P}_{< #1}\,}
\newcommand\true{\mathbf{t}}
\newcommand\false{\mathbf{f}}

\newcommand{\sem}[1][{\ARG}]{\ensuremath{\llbracket #1 \rrbracket}}
\newcommand{\Pf}{F} 

\newcommand\Formn{\mathrm{Fm}_n}
\newcommand\Modn{\mathrm{Mod}_n}
\DeclareMathOperator{\Typ}{Typ}

\newcommand{\PR}{\mathbf{Pries}} 
\newcommand{\BS}{\mathbf{BStone}} 
\newcommand{\DL}{\mathbf{DLat}} 
\newcommand{\BA}{\mathbf{BA}} 

\renewcommand{\P}{\raisebox{.17\baselineskip}{\Large\ensuremath{\wp}}} 
\newcommand{\op}{\mathrm{op}} 
\DeclareMathOperator{\Mea}{\mathcal{M}_{\Gamma}}
\DeclareMathOperator{\Meac}{\mathcal{M}_I}
\newcommand\Shat{\w{\mathbf S}}
\DeclareMathOperator{\B}{\mathcal{B}} 

\DeclareMathOperator{\PFG}{\mathbf P} 
\renewcommand{\restriction}{\mathord{\upharpoonright}}
\newcommand{\blank}{(\ARG)}
\newcommand{\LG}{\gamma^{\#}}    
\newcommand{\LC}{\ic^{\#}}    
\newcommand\PL[1]{P\mkern-1mu\mathcal{L}_{#1}}

\newcommand{\mono}{\hookrightarrow} 
\newcommand{\epi}{\twoheadrightarrow} 
\renewcommand{\epsilon}{\varepsilon}
\renewcommand{\theta}{\vartheta}
\renewcommand{\phi}{\varphi}

\begin{document}

\title[A duality theoretic view on limits of finite structures]{A duality theoretic view on limits of finite structures}
\titlecomment{This is an extended version of a paper with the same title which has appeared in the proceedings of the International Conference on Foundations of Software Science and Computation Structures (FoSSaCS), 299--318 (2020).}

\author[M.~Gehrke]{Mai Gehrke\rsuper{a}}	
\address{CNRS and Universit\'e C{\^o}te d'Azur, Nice, France}
\email{mgehrke@unice.fr}  
\thanks{Mai Gehrke acknowledges support from the European Research Council (ERC) under the European Union's Horizon 2020 research and innovation program (grant agreement No.670624).}	

\author[T.~Jakl]{Tom\'{a}\v{s} Jakl\rsuper{b}}	
\address{Department of Computer Science and Technology, University of Cambridge, UK}
\email{tj330@cam.ac.uk}  	
\thanks{Tom\'a\v s Jakl has received partial support from the EPSRC grant EP/T007257/1.}	

\author[L.~Reggio]{Luca Reggio\rsuper{c}}	
\address{Department of Computer Science, University of Oxford, UK}	
\email{luca.reggio@cs.ox.ac.uk}  	
\thanks{Luca Reggio has received funding from the European Union's Horizon 2020 research and innovation programme under the Marie Sk{\l}odowska-Curie grant agreement No.837724.}	



\begin{abstract}
  \noindent     A systematic theory of \emph{structural limits} for finite models has been developed by Ne{\v s}et{\v r}il and Ossona de Mendez. It is based on the insight that the collection of finite structures can be embedded, via a map they call the \emph{Stone pairing}, in a space of measures, where the desired limits can be computed. We show that a closely related but finer grained space of (finitely additive) measures arises---via Stone-Priestley duality and the notion of types from model theory---by enriching the expressive power of first-order logic with certain ``probabilistic operators''.  We provide a sound and complete calculus for this extended logic and expose the functorial nature of this construction.
    
The consequences are two-fold. On the one hand, we identify the logical gist of the theory of structural limits. On the other hand, our construction shows that the duality theoretic variant of the Stone pairing captures the adding of a layer of quantifiers, thus making a strong link to recent work on semiring quantifiers in logic on words. In the process, we identify the model theoretic notion of \emph{types} as the unifying concept behind this link. These results contribute to bridging the strands of logic in computer science which focus on semantics and on more algorithmic and complexity related areas, respectively.
\end{abstract}

\maketitle
    
\section{Introduction}
While topology plays an important role, via Stone duality, in many parts of semantics, topological methods in more algorithmic and complexity oriented areas of theoretical computer science are not so common. One of the few examples,
 the one we want to consider here, is the study of limits of finite relational structures. 
We will focus on the \emph{structural limits} introduced by Ne{\v s}et{\v r}il and Ossona de Mendez~\cite{nevsetril2012model,NO2020mem}. These provide a common generalisation of various notions of limits of finite structures studied in probability theory, random graphs, structural graph theory, and finite model theory. 
The basic construction in this work is the so-called \emph{Stone pairing}. 
Given a relational signature $\sigma$ and a first-order formula $\phi$ in the signature $\sigma$ with free variables $v_1, \dots, v_n$, define

\begin{equation}\label{eq:Stone-Pairing}
    \SPp{\phi, A} \coloneqq \frac{|\{ \overline a \in A^n \mid A \models \phi(\overline a)\}|}{|A|^n} \qquad \parbox{13em}{\centering\emph{(the probability that a random assignment in $A$ satisfies $\phi$)}.} 
\end{equation}

Ne{\v s}et{\v r}il and Ossona de Mendez view the map $A \mapsto \SPp{\ARG, A}$ as an embedding of the finite $\sigma$-structures into the space of probability measures over the Stone space dual to the Lindenbaum--Tarski algebra of all first-order formulas in the signature $\sigma$. This space is complete and thus provides the desired limit objects for all sequences of finite structures which embed as Cauchy sequences.

Another example of topological methods in an algorithmically oriented area of computer science is the use of profinite monoids in automata theory. In this setting, profinite monoids are the subject of the extensive theory, based on theorems by Eilenberg and Reiterman, and used, among others, to settle decidability questions~\cite{Pin09}.
In~\cite{GGP2008}, it was shown that this theory may be understood as an application of Stone duality, thus making a bridge between semantics and more algorithmically oriented work. 
Bridging this semantics-versus-algorithmics gap in theoretical computer science has since gained quite some momentum, notably with the recent strand of research by Abramsky, Dawar and co-workers~\cite{Abramsky2017b,AbramskyShah2018}. 
In this spirit, a natural question is whether the structural limits of \NOdM{} also can be understood semantically, and in particular whether the topological component may be seen as an application of Stone duality. 

 More precisely, recent work on understanding quantifiers in the setting of logic on finite words~\cite{GPR2017} has shown that adding a layer of certain quantifiers (such as classical and modular quantifiers) corresponds dually to a space-of-measures construction. The measures involved are not classical but only finitely additive and they take values in finite semirings rather than in the unit interval. Nevertheless, this appearance of \emph{measures as duals of quantifiers} begs the further question whether the spaces of measures in the theory of structural limits may be obtained via Stone duality from a semantic addition of certain quantifiers to classical first-order logic.

The purpose of this paper is to address this question. Our main result is that the Stone pairing of \NOdM{} is related by a retraction to a Stone space of finitely additive measures, which is dual to the Lindenbaum--Tarski algebra of a logic fragment obtained from first-order logic by adding one layer of probabilistic quantifiers, and which arises in exactly the same way as the spaces of semiring-valued measures in logic on words. That is, the Stone pairing, although originating from other considerations, may be seen as arising by duality from a semantic construction.

A foreseeable hurdle is that spaces of measures are not zero-dimensional and hence outside the scope of Stone duality. In fact, this issue stems from non-zero dimensionality of the unit interval $\ui$, in which the classical measures are valued. This is well-known to cause problems \eg \ in attempts to combine non-determinism and probability in domain theory~\cite{Jung13}. However, in the structural limits of \NOdM, at the base, one only needs to talk about finite models equipped with normal distributions and thus only the finite intervals $\Lh_n \coloneqq \{ 0, \frac{1}{n}, \frac{2}{n}, \dots, 1\}$ are involved. A careful duality theoretic analysis identifies a codirected diagram (\ie, an inverse limit system) based on these intervals compatible with the Stone pairing. The resulting inverse limit, which we denote $\Lh$, is a Priestley space. It comes equipped with an algebra-like structure, which allows us to reformulate many aspects of the theory of structural limits in terms of $\Lh$-valued measures as opposed to $\ui$-valued measures.

Some interesting features of $\Lh$, dictated by the nature of the Stone pairing and the ensuing codirected diagram, are that:
\begin{itemize}
\item $\Lh$ is based on a version of $\ui$ in which the rationals are doubled;
\item $\Lh$ comes with section-retraction maps $\begin{tikzcd}[cramped,sep=1.8em] \ui \rar[hook]{\ic} & \Lh \rar[->>]{\gamma} & \ui\end{tikzcd}$;
\item the map $\ic$ is lower semicontinuous while the map $\gamma$ is continuous.
\end{itemize}
These features are a consequence of general theory and precisely allow us to witness continuous phenomena relative to $\ui$ in the setting of $\Lh$.

Let us stress that in the present paper we deal with \emph{finitely additive} measures, as opposed to $\sigma$-additive ones. In fact, the Stone pairing construction naturally yields finitely additive measures on a Boolean algebra $B$. In turn, by Stone duality (cf.\ Section~\ref{s:duality} below), $B$ can be identified with the algebra of certain open subsets of a space $X$. As \NOdM{} point out in \cite[Fact~1 p.~373]{nevsetvril2016first}, if $B$ is countable, there is a bijection between finitely additive probability measures $B\to [0,1]$ and $\sigma$-additive (regular) probability measures on the space $X$ equipped with the $\sigma$-algebra of measurable subsets generated by~$B$. Thus, in this framework one can equivalently work with \emph{bona fide} measures on $X$ and this is the route that \NOdM{} take. However, in general, the basic notion arising from the study of structural limits is that of finitely additive measure.

\subsection*{Our contribution}
We show that the ambient space of measures for the structural limits of \NOdM{} can be obtained via \emph{``adding a layer of quantifiers''} in a suitable enrichment of first-order logic. The conceptual framework for seeing this is that of \emph{types} from classical model theory.
 More precisely, we will see that a variant of the Stone pairing is a map into a space of measures with values in a Priestley space $\Lh$. Further, we show that this map is in fact the embedding of the finite structures into the space of ($0$-)types of an extension of first-order logic, which we axiomatise.
On the other hand, $\Lh$-valued measures and $\ui$-valued measures are tightly related by a retraction-section pair which allows the transfer of properties.
These results identify the logical gist of the theory of structural limits and provide a new interesting connection between logic on words and the theory of structural limits in finite model theory.

\subsection*{Outline of the paper.}
In Section \ref{s:prel} we briefly recall Stone-Priestley duality, its application in logic via spaces of types, and the particular instance of logic on words (needed only to show the similarity of the constructions featuring in logic on words and in the theory of structural limits). 

In Section \ref{s:Lh} we introduce the Priestley space $\Lh$ with its additional operations, and show that it admits $\ui$ as a retract (the duality theoretic analysis justifying the structure of $\Lh$ is deferred to the appendix). The spaces of $\Lh$-valued measures are introduced in Section \ref{s:spaces-of-measures}, and the retraction of $\Lh$ onto $\ui$ is lifted to the appropriate spaces of measures. 

In Section \ref{s:stone-pairing}, we introduce the $\Lh$-valued Stone pairing and make the link with logic on words. Further, we compare convergence in the space of $\Lh$-valued measures with the one considered by Ne{\v s}et{\v r}il and Ossona de Mendez. In Section \ref{s:logic-of-measures}, we show that constructing the space of $\Lh$-valued measures dually corresponds to enriching the logic with probabilistic operators. 

To avoid overburdening the reader with technical details, we have decided to postpone the derivation of $\Lh$ and its structure until the appendix. This derivation relies on extended Priestley duality and the theory of canonical extensions, which we recall in Appendix~\ref{s:ext-prie-dual-can-ext}. This general theory is then applied in Appendix~\ref{s:apply-ext-prie-dual} to explain how the space $\Lh$ and its algebra-like structure are derived.

\section{Preliminaries}\label{s:prel}
\paragraph*{\em Notation.} Throughout this paper, if $X\xrightarrow{f} Y\xrightarrow{g} Z$ are functions, their composition is denoted $g\cdot f$. For a subset $S\subseteq X$, $f_{\restriction S}\colon S\to Y$ is the obvious restriction. Given any set $T$, $\P(T)$ denotes its power-set. Further, for a poset $P$, $P^\op$ is the poset obtained by turning the order of $P$ upside down.
\subsection{Stone-Priestley duality}\label{s:duality}
In this paper, we will need Stone duality for bounded distributive lattices in the order-topological form due to Priestley \cite{Priestley1970}. It is a powerful and well established tool in the study of propositional logic and semantics of programming languages, see \eg \ \cite{Goldblatt1989,Abramsky91} for major landmarks and~\cite{GJR2020} for an overview of the role of duality methods in logic and computer science. We briefly recall how this duality works.

A \emph{compact ordered space} is a pair $(X,\leq)$ consisting of a compact space $X$ and a partial order $\leq$ on $X$ closed in the product topology of $X\times X$. (Note that such a space is always Hausdorff.) A compact ordered space is a \emph{Priestley space} provided it is \emph{totally order-disconnected}. That is, for all $x,y\in X$ such that $x\not\leq y$, there is a \emph{clopen} (\ie \ simultaneously closed and open) set $C\subseteq X$ that is an up-set for $\leq$ and satisfies $x\in C$ but $y\notin C$. 

Next, we recall the construction of the Priestley space of a distributive lattice\footnote{Throughout, we assume all distributive lattices are bounded, with the bottom and top elements denoted by $0$ and $1$, respectively. Further, lattice homomorphisms are required to preserve these bounds.} $D$.
 A non-empty proper subset $F\subset D$ is a \emph{prime filter} if it is \emph{(i)} upward closed (in the natural order of $D$), 
 \emph{(ii)} closed under finite meets, and \emph{(iii)} if $a\vee b\in F$, either $a\in F$ or $b\in F$. Denote by $X_D$ the set of all prime filters of $D$. By Stone's Prime Filter Theorem, the map
\begin{align*}
\sem\colon D\to \P(X_D), \ \ a\mapsto \sem[a]\coloneqq \{F\in X_D\mid a\in F\}
\end{align*}
is an embedding.
Priestley's insight was that $D$ can be recovered from $X_D$, if the latter is equipped with the inclusion order and the topology generated by the sets of the form $\sem[a]$ and their complements. This makes $X_D$ into a Priestley space---the \emph{dual space} of $D$---and the map $\sem$ is an isomorphism between $D$ and the lattice of clopen up-sets of $X_D$. Conversely, any Priestley space $X$ is the dual space of the lattice of its clopen up-sets. We call the latter the \emph{dual lattice} of $X$. 

This correspondence extends to morphisms: if $h\colon D_1\to D_2$ is a lattice homomorphism, then $h^{-1}\colon X_{D_2}\to X_{D_1}$ is a \emph{morphism of Priestley spaces}, \ie \ a continuous monotone map. In the other direction, if $f\colon X_1\to X_2$ is a morphism of Priestley, then $f^{-1}$ is a lattice homomorphism from the lattice of clopen up-sets of $X_2$ to the lattice of clopen up-sets of~$X_1$.

Priestley duality states that the ensuing functors are quasi-inverse, hence the category $\DL$ of distributive lattices with homomorphisms is dually equivalent to the category $\PR$ of Priestley spaces and continuous monotone maps.
When restricting to Boolean algebras, we recover the celebrated Stone duality between the full subcategory $\BA$ of $\DL$ defined by Boolean algebras, and the full subcategory $\BS$ of $\PR$ on those objects carrying the identity order. These can be identified with the \emph{Boolean (Stone) spaces}, \ie \ compact Hausdorff spaces in which the clopen subsets form a basis for the topology. 
\begin{figure}[htb]
\centering
\begin{tikzcd}[column sep=4em, row sep=3em]
\DL \arrow[yshift=3pt]{r} & \PR^{\mathrm{op}} \arrow[yshift=-3pt]{l} \\
\BA \arrow[hookrightarrow]{u} \arrow[yshift=3pt]{r} & \BS^{\mathrm{op}} \arrow[yshift=-3pt]{l} \arrow[hookrightarrow]{u}
\end{tikzcd}
\caption{Priestley and Stone dualities.}
\end{figure}

\subsection{Stone duality and logic: type spaces}\label{s:type-spaces}
The \emph{theory of types} is an important tool for first-order logic. We briefly recall the concept as it is closely related to, and provides the link between, two otherwise unrelated occurrences of topological methods in theoretical computer science.

Consider a first-order signature $\sigma$ and a first-order theory $T$ in this signature. Assume that we have a countably infinite sequence of distinct variables $(v_1, v_2, v_3, \dots)$. Then, for each $n\in\N$, let $\Formn$ denote the set of first-order formulas whose free variables are among $\overline{v}=(v_1,\dots,v_n)$, and let $\Modn(T)$ denote the class of all pairs $(A,\alpha)$ where $A$ is a model of $T$ and $\alpha\colon \{v_1,\dots,v_n\} \to A$ is an interpretation of $\overline{v}$ in $A$. 

Then the satisfaction relation, $(A,\alpha)\models \varphi(\overline{v})$, is a binary relation from $\Modn(T)$ to $\Formn$. It induces the equivalence relations of elementary equivalence and logical equivalence on these sets, respectively:
\[
\begin{array}{cccl}
    (A,\alpha)\equiv(B,\beta) &\defiff& \forall \varphi(\overline{v})\in\Formn,  & (A,\alpha)\models \varphi(\overline{v}) \text{ iff } (B,\beta)\models \varphi(\overline{v}),
    \\
    \varphi(\overline{v})\approx\psi(\overline{v}) &\defiff& \forall (A,\alpha)\in\Modn(T), & (A,\alpha)\models \varphi(\overline{v}) \text{ iff } (A,\alpha)\models \psi(\overline{v}).
\end{array}
\]

The quotient $\FO_n(T)\coloneqq \Formn/{\approx}$ carries a natural Boolean algebra structure and is known as the \emph{$n$-th Lindenbaum--Tarski algebra} of $T$. Further, denoting by $[(A,\alpha)]$ the $\equiv$-equivalence class of $(A,\alpha)$, $\Typ_n(T)\coloneqq \Modn(T)/{\equiv}$ is naturally endowed with a topology, generated by the sets $\{ [(A,\alpha)]\mid (A,\alpha)\models \varphi(\overline{v})\}$ for $\varphi(\overline{v})\in\Formn$, and is known as the \emph{space of $n$-types} of $T$. Furthermore, the Lindenbaum--Tarski algebra is often defined as the quotient of $\Formn$ with respect to provable equivalence, where $\phi(\overline{v})$ and $\psi(\overline{v})$ are provably equivalent (modulo $T$) if $T\vdash \phi(\overline{v})\leftrightarrow \psi(\overline{v})$. By G\"odel's completeness theorem, $\FO_n(T)$ is isomorphic to the Lindenbaum--Tarski algebra or, equivalently, $\Typ_n(T)$ is the Stone dual of $\Formn$ modulo provable equivalence. See \eg~\cite[\S 6.3]{Hodges1993}.

The Boolean algebra $\FO(T)$ of \emph{all} first-order formulas modulo logical equivalence over $T$ is the directed colimit of the $\FO_n(T)$ for $n\in\N$ while its dual space, $\Typ(T)$, is the codirected limit of the $\Typ_n(T)$ for $n\in\N$ and consists of (equivalence classes of) models equipped with interpretations of the full set of~variables.

Now, consider the set $\Fin(T)$ consisting of isomorphism classes of finite $T$-models. Because two finite models are elementarily equivalent precisely when they are isomorphic, $\Fin(T)$ can be identified with a subset of the space $\Typ_0(T)$ of  $0$-types of $T$.
If we want to study finite models, there are two equivalent approaches: \eg \ at the level of sentences, we can either consider the theory $T_\text{\em fin}$ of finite $T$-models, or the topological closure of $\Fin(T)$ in the space $\Typ_0(T)$. This closure yields a space, which should tell us about finite $T$-structures. Indeed, it is equal to $\Typ_0(T_\text{\em fin})$, the space of pseudofinite $T$-structures. For an application of this, see~\cite{vanGoolSteinberg}.
Below, we will see an application in finite model theory of the case $T=\emptyset$ (in this case we write $\FO(\sigma)$ and $\Typ(\sigma)$ instead of $\FO(\emptyset)$ and $\Typ(\emptyset)$).

In light of the theory of types as exposed above, the Stone pairing of \NOdM{} (equation~\eqref{eq:Stone-Pairing}) can be seen as an embedding of finite structures into the space of finitely additive probability measures on $\Typ(\sigma)$, which set-theoretically are finitely additive functions $\FO(\sigma) \to \ui$. In fact, we will see that a finer grained variant of the Stone pairing corresponds precisely to embedding finite structures into the space of $0$-types of an extension of first-order logic, \cf \ Theorem~\ref{t:axiomatisation-Boolean-case} and the discussion that follows the theorem.

\subsection{Duality and logic on words}\label{s:duality-and-low}
As mentioned in the introduction, spaces of measures arise via duality in the study of semiring quantifiers in \emph{logic on words}~\cite{GPR2017}. In Section~\ref{s:Gamma-valued-Stone-pairing}, we will show that basically the same construction yields the Stone pairing of \NOdM. To make this analogy precise, here we provide a brief overview of logic on words and the role of measure spaces. As the technical development of the paper is independent from this material, the reader can safely defer the reading of this section.

Logic on words, as introduced by B\"uchi, see \eg \ \cite{MaSch08} for a recent survey, is a variation and specialisation of finite model theory where only models based on words are considered. \emph{I.e.}, a word $w\in A^*$ is seen as a relational structure on $\{1,\ldots, |w|\}$, where $|w|$ is the length of $w$, equipped with a unary relation $P_a$ for each $a\in A$, singling out the positions in $w$ where the letter $a$ appears, and the binary predicate $<$ interpreted as the linear order $1<\cdots<|w|$. Each sentence $\phi$ in a signature interpretable over these structures yields a language $L_{\phi}\subseteq A^*$ consisting of the words satisfying $\phi$. Thus, logic fragments are considered modulo the theory of finite words and the Lindenbaum--Tarski algebras are subalgebras of $\P(A^*)$ consisting of the appropriate $L_{\phi}$'s, \cf \ \cite{vanGoolSteinberg} for a treatment of first-order logic on words. 
 
For lack of logical completeness, the duals of the Lindenbaum--Tarski algebras have more points than those given by models. Nevertheless, the dual spaces of types, which act as compactifications and completions of the collections of models, provide a powerful tool for studying logic fragments by topological means. The central notion is that of \emph{recognition}, in which a Boolean subalgebra $\B\subseteq\P(A^*)$ is studied by means of the dual map $\eta\colon\beta(A^*)\twoheadrightarrow X_{\B}$. Here $\beta(A^*)$ is the Stone dual of $\P(A^*)$, also known in topology as the {\v C}ech-Stone compactification of the discrete space $A^*$, and $X_{\B}$ is the Stone dual of $\B$. The set $A^*$ embeds in $\beta(A^*)$, and $\eta$ is uniquely determined by its restriction $\eta_0\colon A^*\to X_{\B}$. Now, Stone duality implies that $L\subseteq A^*$ is in $\B$ precisely when there is a clopen subset $V\subseteq X_{\B}$ so that 
$\eta_0^{-1}(V)=L$. Anytime the latter is true for a map $\eta$ and a language $L$ as above, one says that\footnote{Here, being beyond the scope of this paper, we are ignoring the important role of the monoid structure available on the spaces (in the form of profinite monoids or BiMs, \cf \ \cite{vanGoolSteinberg,GPR2017}).} \emph{$\eta$ recognises $L$}.

When studying logic fragments via recognition, the following inductive step is central: given a notion of quantifier and a recogniser for a Boolean algebra of formulas with a free variable, construct a recogniser for the Boolean algebra generated by the formulas obtained by applying the quantifier. 
 This problem was solved in \cite{GPR2017}, using duality theory, in a general setting of \emph{semiring quantifiers}. The latter are defined as follows: let $(S,+,\cdot,0_S,1_S)$ be a semiring, and $k\in S$. Given a formula $\psi(v)$, the formula $\exists_{S,k}v.\psi(v)$ is true of a word $w\in A^*$ if, and only if, 
 \[
  \underbrace{1_S+\cdots+ 1_S}_{\text{$m$ times}} = k
 \]
where $m$ is the number of assignments of the variable $v$ in $w$ satisfying $\psi(v)$. 
If $S=\Z/q\Z$, we obtain the so-called \emph{modular quantifiers}, and for $S$ the two-element lattice we recover the existential quantifier $\exists$. 

To deal with formulas with a free variable, one considers maps of the form 
\[
f\colon \beta((A\times 2)^*)\to X,
\] 
where the extra bit in $A\times 2$ is used to mark the interpretation of the free variable. In \cite{GPR2017} (see also \cite{GPR2019}), it was shown that $L_{\psi(v)}$ is recognised by $f$ if, and only if, for every $k\in S$ the language $L_{\exists_{S,k}v.\psi(v)}$ is recognised by the composite
\begin{equation} \label{def:R}
\xi\colon A^* \xrightarrow{\mathmakebox[3em]{R}} \Shat(\beta((A\times 2)^*)) \xrightarrow{\mathmakebox[3em]{\Shat(f)}} \Shat(X),
\end{equation}
 where $\Shat(X)$ is the space of finitely additive $S$-valued measures on $X$,
and $R$ maps $w\in A^*$ to the measure $\mu_w\colon \P((A\times 2)^*) \to S$ sending $K\subseteq (A\times 2)^*$ to the sum $1_S + \cdots + 1_S$, $n_{w,K}$ times. Here, $n_{w,K}$ is the number of interpretations $\alpha$ of the free variable $v$ in $w$ such that the pair $(w,\alpha)$, seen as an element of $(A\times 2)^*$, belongs to $K$. Finally, $\Shat(f)$ sends a measure to its pushforward along $f$.

\section{\texorpdfstring{The space $\Lh$}{The space Gamma}}\label{s:Lh}

\NOdM{} observed in \cite{nevsetril2012model} that the Stone pairing $\SPp{\ARG,\ARG}$ (\cf{} equation~\eqref{eq:Stone-Pairing}), with its second argument fixed, yields a finitely additive function $\FO(\sigma) \to \ui$. In other words, given two first-order formulas $\phi$ and $\psi$ and a finite $\sigma$-structure $A$,
\[ \SPp{\phi, A} + \SPp{\psi,A} = \SPp{\phi\vee \psi,A} + \SPp{\phi\mee \psi, A} \]
and also $\SPp{\false, A} = 0$ and $\SPp{\true, A} = 1$, where $\false$ and $\true$ denote false and true, respectively. By Stone duality, we can identify $\FO(\sigma)$ with the Boolean algebra of clopen subsets of the space of types $\Typ(\sigma)$. By assuming that clopen (equivalently, first-order definable) subsets of $\Typ(\sigma)$ form the Boolean algebra of measurable subsets, we view $\Typ(\sigma)$ as a measurable space. Then, assigning to the second argument of the Stone pairing yields an embedding of $\Fin(\sigma)$, the set of isomorphism classes of finite $\sigma$-structures, into the space $\Meac(\Typ(\sigma))$ of finitely additive measures:
\[ \Fin(\sigma) \to \Meac(\Typ(\sigma)),\qquad A \mapsto \SPp{\ARG, A}. \]

The main obstacle in being able to use duality theory to investigate the logical nature of the space $\Meac(\Typ(\sigma))$ is the fact that this space is not zero-dimensional, and thus not amenable to the usual duality theoretic methods. This is a direct consequence of the fact that $\Meac(\Typ(\sigma))$ is a closed subspace of the product $\ui^{\FO(\sigma)}$, and thus inherits most of its topological properties from the unit interval $\ui$. Consequently, in order to exploit duality theory, \eg \ in the form of Priestley duality, we need to find a suitable Priestley space $\Lh$ that will play the role of the unit interval.

The construction of $\Lh$ comes from the insight that the range of the Stone pairing $\left<\ARG, A\right>$, for a finite $\sigma$-structure $A$ and formulas restricted to a fixed number of free variables, can be confined to a chain of the form
\[ \Lh_n \coloneqq \left\{ 0, \frac{1}{n}, \frac{2}{n}, \dots, 1\right\}, \]
where $n \coloneqq |A|^k$ if $k$ is the number of free variables. Each $\Lh_n$, with the obvious total order and the discrete topology, is a Priestley space. To allow for formulas with any number of free variables, we require that $\Lh$ is obtained as a codirected limit of all finite chains $\Lh_n$. If the maps in the codirected diagram, say $\Lh_{m} \to \Lh_n$, are monotone, we are guaranteed that $\Lh$ is a Priestley space, see \eg \ \cite[Corollary VI.3.3]{Johnstone1986}. There are two natural candidates for such maps, namely the floor and ceiling functions
\[ \lfloor \ARG \rfloor\colon \Lh_{nm} \twoheadrightarrow \Lh_n \qtq{and} \lceil \ARG \rceil\colon \Lh_{nm} \twoheadrightarrow \Lh_n \]
which send $\frac{a}{nm}$ to $\frac{b}{n}$ where  $b \coloneqq \lfloor a/m\rfloor$ and $b \coloneqq \lceil a/m\rceil$, respectively.

Lastly, we require that $\Lh$ comes equipped with some algebraic structure which is rich enough so that we can define $\Lh$-valued measures. Since $\Lh$ will be computed as a limit of the finite chains $\Lh_n$, the natural way to identify this algebraic structure is to analyse the arithmetic operations available on those finite chains. Given that we need to represent finite additivity, it is natural to consider partial addition\footnote{Measures valued in partial monoids have been considered in several places in the literature, see \eg{} \cite{FB1994}.} $x+y$ on $\Lh_n$, defined whenever $x+y \leq 1$. The partial addition on $\Lh_n$ behaves well in many respects, and even has a right adjoint:
\[ x + y \leq z \iff x \leq z - y \]
 where $z - y$ is the partial subtraction defined whenever $y \leq z$. 
However, partial addition is not compatible with floor and ceiling functions. Indeed, the operator dual to the partial addition on chains $\Lh_n$ is not preserved by the lattice homomorphisms dual to the floor functions $\lfloor \ARG \rfloor\colon \Lh_{nm} \to \Lh_n$, nor by the duals of the ceiling functions $\lceil \ARG \rceil\colon \Lh_{nm} \to \Lh_n$. For more details, see Appendix~\ref{s:apply-ext-prie-dual}.

On the other hand, as we shall now see, the partial subtraction is compatible with floor functions.  Note that, by the adjointness property stated above, addition and subtraction are interdefinable, and so we preserve our ability to express finite additivity. Let $L_n$ be the lattice dual to the Priestley space $\Lh_n$, \ie \ the lattice of up-sets of $\Lh_n$. In other words, $L_n$ is the finite chain 
\[ L_n \ee{\cong} \left(\bot < 1 < \frac{n-1}{n} < \dots < \frac{2}{n} < \frac{1}{n} < 0\right). \]
The following lemma provides a description of the operator on $L_n$ dual to partial subtraction on $\Lh_n$, and relies on \emph{extended} Priestley duality between lattices with operators and Priestley spaces with partial operations (or relations), see \eg\ \cite{Goldblatt1989} and also \cite{GP2007,GP2007b}. The necessary background on extended Priestley duality, and a derivation of the following fact, are offered in Appendix~\ref{s:apply-ext-prie-dual}.

\begin{lem}
    The operator $\oplus \colon L_n \times L_n \to L_n$, defined by
    \[ \frac{a}{n} \oplus \frac{b}{n} =
            \begin{cases}
                \bot & \text{ if } a + b > n \\
                \frac{a+b}{n} & \text{ otherwise}
            \end{cases}
       \quad\qtq{and}\quad
       \bot \oplus u = u \oplus \bot = \bot
    \]
    is dual to the partial subtraction $-\colon \Lh_n \times \Lh_n \rightharpoonup \Lh_n$ with domain $\dom(-) = \{ (z,x) \mid x \leq z\}$.
\end{lem}

It is immediate to see that the lattice homomorphisms dual to floor functions are the lattice embeddings $i_{n,nm}\colon L_n \mono L_{nm}$. Moreover, these embeddings preserve $\oplus$, \ie
\[ i_{n,nm}(u \oplus v) = i_{n,nm}(u) \oplus i_{n,nm}(v)\]
for all $u,v\in L_n$. Note that this is not true for the duals of ceiling functions (see Appendix~\ref{s:apply-ext-prie-dual} for details).

\subsection{\texorpdfstring{A concrete representation of $\Lh$}{A concrete representation of Gamma}}
In view of the discussion above, we define $\Lh$ as the limit of the codirected diagram
\[
\left\{\lfloor \ARG \rfloor\colon \Lh_{mn}\twoheadrightarrow \Lh_n\mid m,n\in \N\right\},
\]
where $\N$ carries the opposite of the divisibility order. Since this is a codirected diagram of finite discrete posets and monotone maps, $\Lh$ is naturally equipped with a structure of Priestley space, see \eg \ \cite[Corollary VI.3.3]{Johnstone1986}. The most lucid description of $\Lh$ is obtained by dualising this construction. Because the category of Priestley spaces is dual to the category of distributive lattices, $\Lh$ is dual to the lattice obtained as the colimit $\Lo$ of the directed system of lattice embeddings
\[
\left\{i_{n,nm}\colon L_n \mono L_{nm} \mid m,n\in \N\right\}.
\]
The lattice $\Lo$ is given by
\[ \Lo = \left\{ \bot\right\} \cup \left(\ui\cap\Q\right)^\op , \]
with $\bot <_\Lo q$ and $q \leq_\Lo p$ for every $p \leq q$ in $\ui\cap\Q$. We can then identify the points of $\Lh$ with the prime filters of $\Lo$. Each $q\in \ui\cap\Q$ yields a prime filter $q\cc\coloneqq\{p\in\Lo \mid q\leq p\}$, and each $r\in (0,1]$ yields a prime filter $r\mm\coloneqq\{p\in\Lo\mid r<p\}$. It is not difficult to see that all prime filters of $\Lo$ are of one of these types, and so $\Lh$ can be represented as based on the set 
\[
\left\{ r\mm \mid r\in (0,1]\right\} \cup \left\{ q\cc \mid q\in \ui\cap\Q \right\}.
\] 
The order of $\Lh$ is the unique total order that has $0\cc$ as bottom element, satisfies $r^*< s^*$ if and only if $r<s$ for ${\scriptstyle \ast}\in\{{\scriptstyle-},{\scriptstyle\circ}\}$, and such that $q\cc$ is a cover of $q\mm$ for every rational $q\in(0,1]$ (\ie \ $q\mm<q\cc$, and there is no element strictly in between). In a sense, the elements of the form $q\cc$ are regarded as \emph{exact values}, and those of the form $r\mm$ as \emph{(lower) approximations}. Cf.~Figure~\ref{f:Lo-Lh}.
The topology of $\Lh$ is generated by the sets of the form
\[ 
\upset p\cc \coloneqq \{x\in\Lh \mid p\cc\leq x\}
   \qtq{and}
   \downset q\mm \coloneqq \{x\in\Lh \mid x\leq q\mm\}
\]
for $p,q\in \Q\cap\ui$ such that $q\neq 0$.

\begin{figure}[htb]
\centering
\begin{tikzpicture}
    \begin{scope}
        \node at (0,0) (bot) {};
        \node at (0,0.15) (1) {};
        \node at (0,2) (0) {};

        \draw[densely dotted] (1.center) -- (0.center);

        \node at ($(bot)+(-0.35,0)$) {$\bot$};
        \node at ($(1)+(0.3,0.0)$) {1};
        \node at ($(0)+(0.3,0)$) {0};

        \foreach \pt in {1,0,bot} {
            \draw ($(\pt)-(0.1,0)$) -- ($(\pt)+(0.1,0)$);
        }

        \node at (-1, 1) {$\Lo =$};

        \draw [<->,
        line join=round,
        decorate,
        decoration={
            zigzag,
            segment length=5,
            amplitude=1,
            post=lineto,
            post length=4pt,
            pre length=4pt
        }]  (-2.7,1) to (-4.7,1);
    \end{scope}

    \begin{scope}[xshift=-17em]
        \node at (0,2.15) (1cc) {};
        \node at (0,2) (1mm) {};
        \node at (0,0) (0cc) {};
        \node at (0,1.4) (r) {};
        \node at (0,0.95) (qc) {};
        \node at (0,0.80) (qm) {};
        \node at ($(r) -(0.4,0.0)$) (rmm) {$r\mm$};
        \node at ($(qc)+(0.4,0.05)$) (qcc) {$q\cc$};
        \node at ($(qm)-(0.4,0.05)$) (qmm) {$q\mm$};
        \node at ($(1cc)+(0.4,0)$) {$1\cc$};
        \node at ($(1mm)-(0.4,0.05)$) {$1\mm$};
        \node at ($(0cc)+(0.4,0)$) {$0\cc$};

        \draw[densely dotted] (1mm.center) -- (qc.center);
        \draw[densely dotted] (qm.center) -- (0cc.center);

        \foreach \pt in {1cc,1mm,0cc,r,qm,qc} {
            \draw ($(\pt)-(0.1,0)$) -- ($(\pt)+(0.1,0)$);
        }
        \node at (-1.3, 1) {$\Lh =$};
    \end{scope}
\end{tikzpicture}
\caption{The Priestley space $\Lh$ and its dual lattice $\Lo$.}
\label{f:Lo-Lh}
\end{figure}

\subsection{\texorpdfstring{The algebraic structure on $\Lh$}{The algebraic structure on Gamma}}\label{subs:mip-and-miss}
Because the embeddings $i_{n,nm}\colon L_n \mono L_{nm}$ preserve the operators $\oplus\colon L_n\times L_n \to L_n$, the colimit lattice $\Lo$ is naturally equipped with the operator $\oplus\colon \Lo \times \Lo \to \Lo$ obtained by gluing together all the operators on the finite sublattices $L_n$. By extended Priestley duality, the operator $\oplus$ on $\Lo$ is dual to a pair of partial operations\footnote{Note that the finite chains $\Lh_n$ come equipped, not only with the partial subtraction $\mip$ as explained above, but also with a second partial operation $\miss$. However, this additional piece of information is not needed in order to recover $\oplus$ on $L_n$. This is a specific feature of finite lattices.} $\mip$ and $\miss$ on $\Lh$. The first one, $\mip\colon \domM\to \Lh$, has domain $\domM = \{(x,y)\in \Lh\times \Lh\mid y\leq x\}$ and is defined by 
\[
    \begin{aligned}
        r\cc \mip s\cc &\ee= (r - s)\cc \\
        r\mm \mip s\cc &\ee= (r-s)\mm
    \end{aligned}
    \qquad\qquad
    \begin{rcases*}
        r\cc \mip s\mm \\
        r\mm \mip s\mm
    \end{rcases*}
    \ee=
    \begin{cases}
        (r - s)\cc & \text{ if } r-s \in \Q \\
        (r - s)\mm & \text{ otherwise}.
    \end{cases}
\]
The second partial operation is definable in terms of $\mip$ and is given by
\begin{equation*}
\miss\colon\domM\to \Lh, \ \ x \miss y \coloneqq \bigvee{\left\{ x \mip q\cc \mid y < q\cc \leq x\right\}}.
\end{equation*}
Explicitly, we have $x\miss x=0\cc$ and, whenever $y<x$,
\begin{align*}
    &\begin{rcases*}
        r\cc \miss s\cc \\
        r\mm \miss s\mm \\
        r\mm \miss s\cc
    \end{rcases*}
    \ee= (r-s)\mm,
   &~ r\cc \miss s\mm \ee=
   \begin{cases}
       (r - s)\cc & \text{ if } r-s \in \Q \\
       (r - s)\mm & \text{ otherwise}.
   \end{cases}
   \end{align*}
The proof of the correspondence between $\oplus$ and the pair $(\mip,\miss)$ is included in Appendix~\ref{s:apply-ext-prie-dual}.

The two partial operations on $\Lh$ have specific topological and order-theoretic properties which are crucial in order to be able to reconstruct $\oplus$. Moreover, we need both operations in order to recover the classical $\ui$-valued measures as retracts of $\Lh$-valued measures (\cf \ the proof of Lemma~\ref{l:measure-maps} essential for Theorem~\ref{th:retraction-measures}). 

Here, we collect only the necessary properties of $\mip$ and $\miss$ needed in Section~\ref{s:spaces-of-measures}. Recall that a map between ordered topological spaces is \emph{lower} (resp.\ \emph{upper}) \emph{semicontinuous} provided the preimage of any open down-set (resp.\ open up-set) is open. The next lemma shows, in particular, that $\mip$ and $\miss$ are semicontinuous, a key observation which will allows us to show that the space of $\Lh$-valued measures is a Priestley space (Proposition~\ref{p:measures-Priestley}).
\begin{lem}\label{l:properties-of-mip}\label{l:properties-of-mis}
If $\domM$ is regarded as an ordered subspace of $\Lh \times \Lh{}^\op$, the following hold:
    \begin{enumerate}
        \item $\domM$ is a closed up-set in $\Lh \times \Lh{}^\op$;
        \item $\mip\colon \domM\to \Lh$ and $\miss\colon \domM\to \Lh$ are monotone, \ie \ if $(x,y)\in\domM$, $x\leq x'$ and $y'\leq y$, then $x\mip y\leq x'\mip y'$ and $x\miss y\leq x'\miss y'$;
        \item $\mip\colon \domM\to \Lh$ is lower semicontinuous;
        \item $\miss\colon \domM\to \Lh$ is upper semicontinuous.
    \end{enumerate}
\end{lem}
\begin{proof}
    Items~1--3 as stated above follow from the general theory of \cite{GP2007,GP2007b} and a direct proof is included in Appendix~\ref{s:ext-prie-dual-can-ext}. Item~4 requires extra care because we extended the natural domain of definition of $\miss$ for our application. 
    
    The general theory entails that the partial map $\miss'$ defined as the restriction of $\miss$ to $\{(x,y) \mid y < x\}$ is upper semicontinuous and also that $\dom(\miss')$ is an open up-set in $\Lh\times \Lh^\op$.
    To deduce item~4, let $q\in (0,1]\cap\Q$. Observe that $x = y$ implies that $x \miss y=0\cc \notin \upset q\cc$. Consequently, the preimage of $\upset q\cc$ under $\miss$ is equal to $(\miss')\inv(\upset q\cc)$, which is open in the subspace topology of $\dom(\miss')$. Since $\dom(\miss')$ is open in $\Lh\times \Lh$, the preimage $(\miss)\inv(\upset q\cc)$ is open in $\Lh \times \Lh$ and thus also in the subspace topology of $\dom(\miss)$.
\end{proof}

\subsection{\texorpdfstring{The retraction $\Lh\epi \ui$}{The retraction Gamma --> [0,1]}}\label{s:retraction-Lh-ui}
In this section we show that, with respect to appropriate topologies, the unit interval $\ui$ can be obtained as a topological retract of $\Lh$, in a way that is compatible with the operation $\mip$. This will be important in Sections~\ref{s:spaces-of-measures} and~\ref{s:stone-pairing}, where we need to move between \ui-valued and $\Lh$-valued measures.
Let us define the monotone surjection given by collapsing the doubled elements:
\begin{equation}\label{eq:gamma}
\gamma\colon \Lh \to \ui, \ \ r\mm, r\cc \mapsto r.
\end{equation}
The map $\gamma$ has a right adjoint, given by
\begin{equation}\label{eq:ic}
\ic\colon \ui \to \Lh, \ r \mapsto
        \begin{cases}
            r\cc & \text{if } r\in \Q \\
            r\mm & \text{otherwise}.
        \end{cases}
\end{equation}
Indeed, it is readily seen that, for all $y\in \Lh$ and $x\in \ui$, $\gamma(y)\leq x$ if and only if $y\leq \ic(x)$. The composition $\gamma\cdot\ic$ coincides with the identity on $\ui$, \ie \ $\ic$ is a section of $\gamma$. Moreover, as a consequence of the next lemma, this retraction lifts to a topological retract provided we equip $\Lh$ and $\ui$ with the topologies consisting of the open down-sets.

\begin{lem}\label{l:gamma-ic-p-morphisms}
    The map $\gamma\colon \Lh \to \ui$ is continuous and the map $\ic\colon \ui \to \Lh$ is lower semicontinuous.
\end{lem}
\begin{proof}
    To check continuity of $\gamma$ observe that, for $q\in\Q \cap (0,1)$, the sets $\gamma\inv(q,1]$ and $\gamma\inv[0,q)$ coincide, respectively, with the open sets
    \[
         \bigcup \left\{ \upset p\cc \mid  p\in \Q\cap\ui \text{ and } q < p \right\} \ete{and}
         \bigcup \left\{ \downset p\mm \mid  p\in \Q\cap (0,1] \text{ and } p < q \right\}.
     \]
    Also, $\ic$ is lower semicontinuous, for $\ic\inv(\downset q\mm) = [0, q)$ whenever $q\in \Q\cap(0,1]$.
\end{proof}

It is easy to see that both $\gamma$ and $\ic$ preserve the minus structure available on $\Lh$ and $\ui$ (the unit interval is equipped with the usual minus operation $x - y$ defined whenever $y\leq x$), that is,
\begin{itemize}
    \item  $\gamma(x \mip y) = \gamma(x \miss y) = \gamma(x) - \gamma(y)$ whenever $y \leq x$ in $\Lh$, and
    \item  $\ic(x - y) = \ic(x) \mip \ic(y)$ whenever $y \leq x$ in $\ui$.
\end{itemize}

\begin{rem}
$\ic\colon \ui\to\Lh$ is not upper semicontinuous because, for every $q\in\Q\cap\ui$,
\[
\ic\inv\left(\upset q\cc\right)=\left\{x\in\ui\mid q\cc \leq \ic(x)\right\}=\left\{x\in \ui\mid \gamma(q\cc)\leq x\right\}=[q,1].
\]
\end{rem}

\begin{rem}\label{rem:im}
The map $\gamma\colon \Lh\to \ui$ has, in addition to the right adjoint $\ic$, also a left adjoint $\im$, given by 
\[
\im\colon \ui \to \Lh, \ x \mapsto
        \begin{cases}
            0\cc & \text{if } x=0 \\
            x\mm & \text{otherwise}.
        \end{cases}
\]
The latter is also a section of $\gamma$, but in the following we shall mainly work with the map $\ic$.
\end{rem}

\section{\texorpdfstring{Spaces of measures valued in $\Lh$ and in $\ui$}{Spaces of measures valued in Gamma and in [0,1]}}\label{s:spaces-of-measures}
The aim of this section is to replace $\ui$-valued measures  by $\Lh$-valued measures. The reason for doing this is two-fold. First, the collection of all $\Lh$-valued measures is a Priestley space (Proposition~\ref{p:measures-Priestley}), and thus amenable to a duality theoretic treatment and a dual logic interpretation (\cf \ Section~\ref{s:logic-of-measures}). Second, it retains more topological information than the space of $\ui$-valued measures. Indeed, the former retracts onto the latter (Theorem~\ref{th:retraction-measures}). 

Let $D$ be a distributive lattice. Recall that, classically, a monotone function 
\[
m\colon D\to \ui
\] 
is a finitely additive probability measure provided it satisfies $m(0) = 0$, $m(1) = 1$, and $m(a) + m(b)=m(a\vee b) + m(a\wedge b)$ for all $a, b\in D$. The latter property is equivalently expressed as
\begin{align}\label{e:finite-additive-with-minuses}
    \forall a,b\in D, \ \ m(a)-m(a\wedge b)=m(a\vee b)-m(b).
\end{align}
We write $\Meac(D)$ for the set of all (finitely additive, probability) measures $D\to\ui$, and regard it as an ordered topological space, with the structure induced by the product order and product topology of $\ui^D$. The notion of (finitely additive, probability) $\Lh$-valued measure is analogous to the classical one, except that the finite additivity property~\eqref{e:finite-additive-with-minuses} splits into two conditions, involving $\mip$ and $\miss$.
\begin{defi}\label{d:measure}
    Let $D$ be a distributive lattice. A \emph{$\Lh$-valued measure} (or simply a \emph{measure}) on $D$ is a function $\mu\colon D\to \Lh$ such that 
    \begin{enumerate}
        \item $\mu$ is monotone,
        \item $\mu(0) = 0\cc$ and $\mu(1) = 1\cc$, and
        \item for all $a,b\in D$,
            \[ \mu(a)\miss \mu(a\wedge b) \leq \mu(a\vee b) \mip \mu(b) \ete{ and } \mu(a) \mip \mu(a\wedge b) \geq \mu(a\vee b) \miss \mu(b).\]
    \end{enumerate}
    We denote by $\Mea(D)$ the subset of $\Lh^D$ consisting of the measures $\mu\colon D\to \Lh$.
\end{defi}
Note that, if we only required one of the two inequalities in item 3 above, we would essentially be considering finitely subadditive measures.

Since $\Lh$ is a Priestley space, so is $\Lh^D$ equipped with the product order and topology. Hence, we regard $\Mea(D)$ as an ordered topological space, whose topology and order are induced by those of $\Lh^D$. 
Explicitly, given measures $\mu,\nu\in\Mea(D)$ we have $\mu\leq \nu$ if, and only if, $\mu(a)\leq\nu(a)$ for all $a\in D$. Moreover, the topology of $\Mea(D)$ is generated by the sets of the form
\[
\MBD{a}{q} \coloneqq \left\{ \nu\in\Mea(D) \mid  \nu(a) < q\cc \right\}
\] 
and 
\[
\MBU{a}{q} \coloneqq \left\{ \nu\in\Mea(D) \mid  \nu(a) \geq q\cc \right\},
\]
for all $a\in D$ and $q\in \Q\cap \ui$. Note that the sets $\MBD{a}{q}$ are clopen down-sets, while the sets \MBU{a}{q} are clopen up-sets.
It turns out that $\Mea(D)$ is a Priestley space (Proposition~\ref{p:measures-Priestley}). The proof of this fact relies on the following lemma:
\begin{lem}\label{l:order-equaliser}
Let $X,Y$ be compact ordered spaces, $f\colon X \to Y$ a lower semicontinuous function, and $g\colon X\to Y$ an upper semicontinuous function. If $X'$ is a closed subset of $X$, then so is $\{ x\in X' \mid  g(x) \leq f(x)\}$.
\end{lem}
\begin{proof}
Whenever $g(x) \not\leq f(x)$ for some $x\in X$, there are an open down-set $U$ and open up-set $V$ of $Y$ that are disjoint and satisfy $f(x) \in U$ and $g(x) \in V$. See \eg \ \cite[Theorem 4 p.\ 46]{Nachbin1965}. Then, the open set $f\inv(U)\cap g\inv(V)$ contains $x$ and is disjoint from $\{x\in X \mid  g(x) \leq f(x)\}$. Since $x$ was arbitrary, we conclude that the set $E\coloneqq \{x\in X \mid  g(x) \leq f(x)\}$ is closed.
Therefore, if $X'$ is a closed subset of $X$, we deduce that $\{ x\in X' \mid  g(x) \leq f(x)\}=E\cap X'$ is also a closed subset of $X$.
\end{proof}
\begin{prop}\label{p:measures-Priestley} 
    For any distributive lattice $D$, $\Mea(D)$ is a Priestley space.
\end{prop}
\begin{proof}
    It suffices to show that $\Mea(D)$ is a closed subspace of $\Lh^D$. Let
\[ 
C_{1,2} \coloneqq \left\{f\in \Lh^D\mid f(0) = 0\cc\right\}\cap \left\{f\in \Lh^D\mid f(1) = 1\cc\right\} \cap \bigcap_{a\leq b} \left\{ f\in \Lh^D \mid f(a) \leq f(b) \right\}. 
\]
Note that the evaluation maps $\ev_a\colon \Lh^D\to \Lh$, $f\mapsto f(a)$ are continuous for every $a\in D$. Thus, the first set in the intersection defining $C_{1,2}$ is closed because it is the equaliser of the evaluation map $\ev_0$ and the constant map of value $0\cc$. Similarly for the set $\{f\in \Lh^D\mid f(1) = 1\cc\}$. The last set is the intersection of sets of the form $\langle \ev_a, \ev_b\rangle\inv (\leq)$, which are closed because $\leq$ is closed in $\Lh\times \Lh$. Hence, $C_{1,2}$ is a closed subset of $\Lh^D$. Moreover,
    \begin{align*}
        \Mea(D) = \bigcap_{a,b\in D} &\left\{ f\in C_{1,2} \mid f(a)\miss f(a\wedge b) \leq f(a\vee b) \mip f(b) \right\} \\
        \cap \bigcap_{a,b\in D} &\left\{ f\in C_{1,2} \mid f(a)\mip f(a\wedge b) \geq f(a\vee b) \miss f(b) \right\}.
    \end{align*}
    From semicontinuity of $\mip$ and $\miss$ (Lemma \ref{l:properties-of-mis}), and Lemma~\ref{l:order-equaliser}, we conclude that $\Mea(D)$ is a closed subspace of $\Lh^D$.
\end{proof}
 It turns out that the construction $D\mapsto \Mea(D)$ is functorial:
 \begin{lem}
 The assignment $D\mapsto \Mea(D)$ can be extended to a contravariant functor $\Mea\colon \DL\to \PR$ by setting, for all lattice homomorphisms $h\colon D\to E$, 
 \[
 \Mea(h)\colon \Mea(E) \to \Mea(D), \ \ \mu\mapsto \mu\cdot h.
 \]
 \end{lem}
 \begin{proof}
  First, we show that if $h\colon D\to E$ is a lattice homomorphism and $\mu\colon E\to \Lh$ is a measure, the composite map $\mu \cdot h\colon D\to \Lh$ is also a measure.
    Since $\mu$ and $h$ are monotone, so is $\mu\cdot h$. Further, $\mu(h(0)) = \mu(0) = 0\cc$, and $\mu(h(1)) = \mu(1) = 1\cc$. For the third condition in Definition~\ref{d:measure} observe that, for all $a, b\in D$,
    \begin{align*}
        \mu(h(a)) \miss \mu(h(a\mee b)) &= \mu(h(a)) \miss \mu(h(a)\mee h(b)) \\
        &\leq \mu(h(a)\vee h(b)) \mip \mu(h(b)) \\
        &= \mu(h(a\vee b)) \mip \mu(h(b)),
    \end{align*}
    where the middle inequality holds because $\mu$ is a measure. The inequality 
    \[
    \mu(h(a)) \mip \mu(h(a\mee b)) \geq \mu(h(a\vee b)) \miss \mu(h(b))
    \] 
    is proved similarly. Thus, $\mu\cdot h \colon D\to \Lh$ is a measure.
    
    The mapping $\Mea(h)\colon \Mea(E) \to \Mea(D)$, $\mu\mapsto \mu\cdot h$, is clearly monotone. With respect to continuity, recall that the topology of $\Mea(D)$ is generated by the sets of the form $\MBD{a}{q} = \{ \nu\in\Mea(D) \mid  \nu(a) < q\cc \}$ and $\MBU{a}{q} = \{ \nu\in\Mea(D) \mid  \nu(a) \geq q\cc \}$,
with $a\in D$ and $q\in \Q\cap \ui$. We have
    \begin{equation*}
        \Mea(h)\inv\left(\MBD{a}{q}\right)=\left\{ \mu\in\Mea(E) \mid  \mu(h(a)) < q\cc \right\} = \MBD{h(a)}{q}
    \end{equation*}
    which is open in $\Mea(E)$. Similarly, $\Mea(h)\inv(\MBU{a}{q})=\MBU{h(a)}{q}$, showing that $\Mea(h)$ is continuous. Hence, in view of Proposition~\ref{p:measures-Priestley}, $\Mea(h)$ is a morphism of Priestley spaces.
It is not difficult to see that identities and compositions are preserved, therefore $\Mea\colon \DL\to \PR$ is a contravariant functor.
 \end{proof}
 \begin{rem}\label{rm:functor-M-cov}
  We work with the contravariant functor $\Mea\colon \DL\to \PR$ because $\Mea$ is concretely defined on the lattice side. However, by Priestley duality, $\DL$ is dually equivalent to $\PR$, so we can think of $\Mea$ as a covariant functor $\PR\to\PR$ (this is the perspective traditionally adopted in analysis, and also in the works of Ne{\v s}et{\v r}il and Ossona de Mendez). From this viewpoint, Section~\ref{s:logic-of-measures} provides a description of the endofunctor on $\DL$ dual to $\Mea\colon \PR\to\PR$.
\end{rem}
Next, we establish a property of the functor $\Mea\colon \DL\to \PR$ which is very useful when approximating a fragment of a logic by smaller fragments (see, \eg, Section \ref{s:Gamma-valued-Stone-pairing}).
\begin{prop}\label{p:functor-mea}
  The contravariant functor $\Mea\colon \DL\to \PR$ sends directed colimits to codirected limits.
\end{prop}
\begin{proof}
Let $D$ be the colimit in $\DL$ of a directed diagram 
\[
\left\{h_{i,j}\colon D_i \to D_j\right\}_{i\leq j},
\] 
with colimit maps $\iota_{i}\colon D_i \to D$ (where $i$ and $j$ vary in a directed poset $I$). We must show that the cone $\{\Mea(\iota_{i})\colon \Mea(D)\to \Mea(D_i)\}_{i\in I}$ is the limit of the diagram
    \[  \left\{
        \begin{tikzcd}[cramped, sep=4em]
            \Mea(D_i) &
            \ar[swap]{l}{\Mea(h_{i,j})}
            \Mea(D_j)
        \end{tikzcd}
        \right\}_{i \leq j}.
    \]
    Let $Y$ be a Priestley space and $\{g_i\colon Y \to \Mea(D_i)\}_{i\in I}$ a set of Priestley morphisms such that $g_i = \Mea(h_{i,j}) \cdot g_j$, for every $i \leq j$. We need to prove that there is a unique Priestley morphism $\xi\colon Y \to \Mea(D)$ such that $\Mea(\iota_i) \cdot \xi = g_i$ for every $i$. 
    
Note that, since directed colimits in $\DL$ are computed in the category of sets, for every element $a$ of $D$ there are an index $i\in I$ and an element $a_i\in D_i$ such that $\iota_i(a_i) = a$. Define
 \[ 
    \xi\colon Y \to \Mea(D),\quad y\mapsto \left(D\xrightarrow{\xi(y)}\Lh, \ a\mapsto g_i(y)(a_i)\right),
 \]
where $a_i$ satisfies $\iota(a_i)=a$. This definition is independent of the choice of $a_i$: for any other $a_j \in D_j$ such that $\iota_j(a_j) = a$, we have that $h_{i,k}(a_i) = h_{j,k}(a_j)$ for some $k \geq i,j$, and so 
\begin{align*}
g_i(y)(a_i) &= g_k(y)(h_{i,k}(a_i)) \\ 
&= g_k(y)(h_{j,k}(a_j)) = g_j(y)(a_j)
\end{align*}
for all $y\in Y$. Observe that, for every $y\in Y$, $\xi(y)$ is indeed a measure $D\to \Lh$. For example, let $a, b\in D$. Then, because $I$ is directed, there is an index $i$ such that $a = \iota_i(a_i)$ and $b = \iota_i(b_i)$ for some $a_i, b_i\in D_i$. Since $g_i(y)$ is a measure, we have that
    \begin{align*}
        \xi(y)(a) \miss \xi(y)(a\mee b)
            &= g_i(y)(\iota_i(a_i)) \miss g_i(y)(\iota_i(a_i)\mee \iota_i(b_i)) \\
            &\leq g_i(y)(\iota_i(a_i)\vee \iota_i(b_i)) \mip g_i(y)(\iota_i(b_i)) \\
            &= \xi(y)(a\vee b) \mip \xi(y)(b).
    \end{align*}
    The other properties are proved in the same spirit. Next, we show that $\xi$ is a Priestley morphism. Monotonicity is immediate: if $y\leq y'$ in $Y$ and $a_i\in D_i$, then 
    \begin{align*}
    \xi(y)(\iota_i(a_i)) &= g_i(y)(a_i) \\
    & \leq g_i(y')(a_i) = \xi(y')(\iota_i(a_i)).
    \end{align*}
    For continuity, for all $a_i \in D_i$ and $q\in \Q\cap \ui$, we have
    \begin{align*}
        \xi\inv\left(\MBD{\iota_i(a_i)}{q}\right)
        = \left\{ y \mid \xi(y)(\iota_i(a_i)) < q\cc \right\}
        = \left\{ y \mid g_i(y)(a_i) < q\cc \right\} = g_i\inv\left(\MBD{a_i}{q}\right)
    \end{align*}
    and, similarly, $\xi\inv(\MBU{\iota_i(a_i)}{q}) = g_i\inv(\MBU{a_i}{q})$. Lastly, we show that $\xi$ is unique with these properties. Let $\lambda\colon Y\to \Mea(D)$ be a Priestley morphism such that $g_i = \Mea(\iota_i)\cdot \lambda$ for every $i\in I$. For every $a\in D$, and element $a_i\in D_i$ such that $\iota_i(a_i)=a$, we get 
    \[
    \lambda(y)(a)=\lambda(y)(\iota_i(a_i)) = g_i(y)(a_i) = \xi(y)(\iota_i(a_i))=\xi(y)(a)
    \]
    for all $y\in Y$. Hence, $\lambda=\xi$.
\end{proof}

Recall the maps $\gamma\colon \Lh\to \ui$ and $\ic\colon \ui\to\Lh$ from equations \eqref{eq:gamma}--\eqref{eq:ic}. In Section~\ref{s:retraction-Lh-ui} we have shown that this is a retraction-section pair. In Theorem \ref{th:retraction-measures} we will lift this retraction to the spaces of measures. We start with an easy observation:
\begin{lem}\label{l:measure-maps}
Let $D$ be a distributive lattice. The following statements hold:
\begin{enumerate}
\item for every $\mu\in\Mea(D)$, $\gamma\cdot\mu\in\Meac(D)$;
\item for every $m\in\Meac(D)$, $\ic\cdot m\in\Mea(D)$.
\end{enumerate}
\end{lem}
\begin{proof}
$(1)$ \ The only non-trivial condition to verify is finite additivity. As pointed out after Lemma \ref{l:gamma-ic-p-morphisms}, the map $\gamma$ preserves both minus operations on $\Lh$. Hence, for all $a,b\in D$, the inequalities 
$\mu(a) \miss \mu(a\wedge b) \leq \mu(a\vee b) \mip \mu(b)$ and
 $\mu(a) \mip \mu(a\wedge b) \geq \mu(a\vee b) \miss \mu(b)$ imply that
$\gamma\cdot\mu(a)-\gamma\cdot\mu(a\wedge b)=\gamma\cdot\mu(a\vee b)-\gamma\cdot\mu(b)$.

$(2)$ \ The first two conditions in Definition~\ref{d:measure} are immediate. The third condition follows from the fact that $\ic(r - s) = \ic(r) \mip \ic(s)$ whenever $s \leq r$ in $\ui$, and $x\miss y\leq x\mip y$ for every $(x,y)\in \domM$.
\end{proof} 

In view of the previous lemma, there are well-defined functions
\[
    \LG\colon \Mea(D) \to \Meac(D), \  \mu \mapsto \gamma\cdot \mu \ee{\text{ and }}
       \LC\colon \Meac(D) \to \Mea(D), \ m \mapsto \ic\cdot m.
\]
\begin{lem}\label{l:gamma-continuous-monotone}
$\LG\colon \Mea(D) \to \Meac(D)$ is a continuous and monotone map.
\end{lem}
\begin{proof}
The topology of $\Meac(D)$ is generated by the sets of the form $\{m\in \Meac(D)\mid m(a)\in O\}$, for $a\in D$ and $O$ an open subset of $\ui$. In turn,
\[
(\LG)\inv \{m\in \Meac(D)\mid m(a)\in O\}=\{\mu\in\Mea(D)\mid \mu(a)\in \gamma\inv (O)\}
\]
is open in $\Mea(D)$ because $\gamma\colon \Lh\to\ui$ is continuous by Lemma \ref{l:gamma-ic-p-morphisms}. This shows that $\LG\colon \Mea(D) \to \Meac(D)$ is continuous. 
Monotonicity is immediate.
\end{proof}

Note that $\LG\colon \Mea(D) \to \Meac(D)$ is surjective, since it admits $\LC$ as a (set-theoretic) section. It follows that $\Meac(D)$ is a compact ordered space:
\begin{cor}\label{c:class-meas-compact-pospace}
For each distributive lattice $D$, $\Meac(D)$ is a compact ordered space.
\end{cor}
\begin{proof}
The surjection $\LG\colon \Mea(D) \to \Meac(D)$ is continuous (Lemma~\ref{l:gamma-continuous-monotone}). Since $\Mea(D)$ is compact by Proposition \ref{p:measures-Priestley}, so is $\Meac(D)$.
The order of $\Meac(D)$ is clearly closed in the product topology, thus $\Meac(D)$ is a compact ordered space.
\end{proof}
Finally, we see that the set-theoretic retraction of $\Mea(D)$ onto $\Meac(D)$ lifts to the topological setting, provided we restrict to the down-set topologies. 
If $(X,\leq)$ is a partially ordered topological space, write $X^{\downarrow}$ for the space with the same underlying set as $X$ and whose topology consists of the open down-sets of $X$.
\begin{thm}\label{th:retraction-measures}
    The maps $\LG\colon \Mea(D)^{\downarrow} \to \Meac(D)^{\downarrow}$ and $\LC\colon \Meac(D)^{\downarrow} \to \Mea(D)^{\downarrow}$ are a retraction-section pair of topological spaces.
\end{thm}
\begin{proof}
It suffices to show that $\LG$ and $\LC$ are continuous. It is not difficult to see, using Lemma \ref{l:gamma-continuous-monotone}, that $\LG\colon \Mea(D)^{\downarrow} \to \Meac(D)^{\downarrow}$ is continuous. For the continuity of $\LC$, note that the topology of $\Mea(D)^{\downarrow}$ is generated by the sets of the form 
$\MBD{a}{q}$, for $a\in D$ and $q\in\Q\cap (0,1]$.
 We have 
\begin{align*}
(\LC)\inv \MBD{a}{q} &=\left\{m\in\Meac(D)\mid m(a)\in \ic\inv(\downset q\mm)\right\} \\
&=\left\{m\in\Meac(D)\mid m(a)<q\right\},
\end{align*}
which is an open set in $\Meac(D)^{\downarrow}$. This concludes the proof.
\end{proof}

\section{\texorpdfstring{The $\Lh$-valued Stone pairing and limits of finite structures}{The Gamma-valued Stone pairing and limits of finite structures}}\label{s:stone-pairing}
In the work of Ne{\v s}et{\v r}il and Ossona de Mendez, the Stone pairing $\left<\ARG,A\right>$ is $\ui$-valued, \ie \ an element of $\Meac(\FO(\sigma))$.
In this section, we show that basically the same construction for the recognisers arising from the application of a layer of semiring quantifiers in logic on words (\cf \ Section~\ref{s:duality-and-low}) provides an embedding of finite $\sigma$-structures into the space of $\Lh$-valued measures. It turns out that this embedding is a $\Lh$-valued version of the Stone pairing. Hereafter we make a notational difference, writing $\SPc{\ARG,\ARG}$ for the (classical) $\ui$-valued Stone pairing. 

The main ingredient of the construction are the $\Lh$-valued finitely supported functions.
To start with, we point out that the partial operation $\mip$ on $\Lh$ uniquely determines---by adjointness---a partial ``plus'' operation on $\Lh$. 
Define
\[ 
+\colon \domP \to \Lh, \qtq{where} \domP = \left\{ (x,y) \mid x \leq 1\cc \mip y\right\} ,
\]
by the following rules (whenever the expressions make sense):
\[
        r\cc+s\cc=(r+s)\cc, \ \ 
        r\mm+s\cc=(r+s)\mm, \ \
        r\cc+s\mm=(r+s)\mm,\ete{and }
        r\mm+s\mm=(r+s)\mm.
\]
Then, for every $x\in \Lh$, the function $\blank + x$ sending $y$ to $y+x$ is left adjoint to the function $\blank \mip x$ sending $y$ to $y\mip x$. This is the content of the next lemma (where, for all $x,y\in \Lh$, we let $[x,y]\coloneqq \{z\in \Lh\mid x\leq z\leq y\}$); for a proof, see the end of Appendix~\ref{s:deriving-operations}.
\begin{lem}\label{lem:mip-is-right-adjoint}
For every $x\in \Lh$, the monotone map $\blank + x\colon [0\cc,1\cc\mip x]\to [x,1\cc]$ is left adjoint to $\blank \mip x\colon [x,1\cc]\to [0\cc,1\cc\mip x]$.
\end{lem}

This plus operation allows us to define the notion of finitely supported $\Lh$-valued function:

\begin{defi}
For any set $X$, $\Fs(X)$ is the set of all functions $f\colon X\to \Lh$ such that
\begin{enumerate}
    \item the set $\supp(f) \coloneqq \{ x\in X \mid   f(x) \not= 0\cc\}$ is finite, and
    \item $f(x_1)+\cdots +f(x_n)$ is defined and equal to $1\cc$, where $\{x_1, \ldots, x_n\}=\supp(f)$.
\end{enumerate}
\end{defi}

To improve readability, if the sum $y_1+\cdots +y_m$ exists in $\Lh$, we denote~it $\sum_{i=1}^{m}{y_i}$. Finitely supported functions in the above sense always determine measures over the power-set algebra:
\begin{lem}\label{l:fs-lift-to-mea}
    Let $X$ be any set. There is a well-defined mapping $\int\colon\Fs(X) \to \Mea(\P(X))$, assigning to every $f\in \Fs(X)$ the measure 
    \[\textstyle
    \int f\colon M\mapsto \int_M f \coloneqq \sum_{x\in M\cap \supp(f)}{f(x)}.
    \]
\end{lem}
\begin{proof}
 We wish to prove that $\int f$ is a measure for every $f\in \Fs(X)$. First, observe that $\int_\emptyset f = 0\cc$ and $\int_X f = 1\cc$ by definition. Moreover, $M \mapsto \int_M f$ is monotone because $x \leq x + y$ for all $x, y\in \Lh$ for which $x + y$ is defined.

    It remains to show that, for any $M, N\sue X$,
    \begin{align} \label{e:meas-eq-for-int}
        \textstyle
        \int_M f \miss \int_{M\cap N} f \,\leq\, \int_{M\cup N} f \mip \int_N f
        \qtq{and}
        \int_M f \mip \int_{M\cap N} f \,\geq\, \int_{M\cup N} f \miss \int_N f.
    \end{align}
    To simplify these expressions, set $m \coloneqq \int_{M\setminus N} f$,  $n \coloneqq \int_{N\setminus M} f$, and $i \coloneqq \int_{M\cap N} f$. Then, we see that
    \[ \textstyle
       \int_M f = m + i,\quad \int_N f = n + i\ete{and} \int_{M\cup N} = m + n + i.
    \]
    Consequently, the inequalities in~\eqref{e:meas-eq-for-int} can be rewritten, respectively, as
    \[
        (m+i) \miss i \,\leq\, (m+n+i) \mip (n+i)
        \qtq{and}
        (m+i) \mip i \,\geq\, (m+n+i) \miss (n+i).
    \]
    By Lemma~\ref{lem:mip-is-right-adjoint}, these are equivalent, respectively, to
    \[
        ((m+i) \miss i) +  (n+i) \,\leq\, m+n+i
        \qtq{and}
        ((m+n+i) \miss (n+i)) + i \leq m+i.
    \]
Next, observe that, for all $x, y\in \Lh$ for which $x + y$ is defined, $(x+y) \miss y$ is always of ${}\mm$-type or equal to $0\cc$, and so $(x+y) \miss y = \im(\gamma(x))$ (where the map $\im\colon \ui\to \Lh$ is defined as in Remark~\ref{rem:im}). Therefore, $((m+i) \miss i) +  n+i = \im(\gamma(m)) + n + i \leq m + n + i$ and $((m+n+i) \miss (n+i)) + i = \im(\gamma(m)) + i \leq m + i$.
\end{proof}

\subsection{\texorpdfstring{The $\Lh$-valued Stone pairing and logic on words}{The Gamma-valued Stone pairing and logic on words}}\label{s:Gamma-valued-Stone-pairing}
Fix a countably infinite sequence of distinct variables $(v_1, v_2, v_3, \dots)$. Recall that $\FO_n(\sigma)$ is the Lindenbaum--Tarski algebra of first-order formulas with free variables among $\{v_1, \dots, v_n\}$. The dual space of $\FO_n(\sigma)$ is the space of $n$-types $\Typ_n(\sigma)$. Its points are the elementary equivalence classes of pairs $(A, \alpha)$, where $A$ is a $\sigma$-structure and $\alpha\colon \{v_1, \dots, v_n\} \to A$ is an interpretation of the variables. Also, recall that $\Fin(\sigma)$ denotes the set of isomorphism classes of finite $\sigma$-structures.

We would like to define a function $\Fin(\sigma) \to \Fs(\Typ_n(\sigma))$ which, intuitively, associates with a finite $\sigma$-structure $A$ the (normalised) characteristic function of the finite set 
\[
\left\{(A,\alpha)\mid \alpha\colon \{v_1, \dots, v_n\} \to A\right\},
\]
which assigns value $\left(\frac{1}{|A|^{n}}\right)\cc$ to $(B,\beta)$ if $A=B$, and $0$ otherwise. However, we need to slightly modify this definition to take into account the fact that elements of $\Typ_n(\sigma)$ are elementary equivalence classes of pairs $(B,\beta)$, and such an equivalence class may contain several pairs whose first component is $A$. 

Define the map $\Fin(\sigma) \to \Fs(\Typ_n(\sigma))$ which sends the isomorphism class of a finite $\sigma$-structure $A$ to the finitely supported function $f_n^A$ whose value at an equivalence class $E \in \Typ_n(\sigma)$ is
\[
    f_n^A(E)\coloneqq \sum_{(A, \alpha) \in E} \left(\frac{1}{|A|^{n}}\right)\cc
    \quad \parbox{21.5em}{\centering\emph{(add $\left(\frac{1}{|A|^n}\right)\cc$ for every interpretation $\alpha$ of the free variables s.t.\ $(A, \alpha)$ is in the equivalence class)}.}
\]
It is not difficult to see that this definition does not depend on the choice of a representative in the isomorphism class of $A$.

By Lemma~\ref{l:fs-lift-to-mea}, we get a measure $\int f_n^A\colon \P(\Typ_n(\sigma))\to \Lh$. Now, for each $\phi\in\FO_n(\sigma)$, let $\sem[\phi]_n \sue \Typ_n(\sigma)$ be the set of (equivalence classes of) $\sigma$-structures with interpretations satisfying $\phi$. By Stone duality we obtain an embedding $\sem_n\colon \FO_n(\sigma)\mono \P(\Typ_n(\sigma))$. Restricting $\int f_n^A$ to $\FO_n(\sigma)$, we get a measure
\[\textstyle 
\mu_n^A \colon \FO_n(\sigma)\to \Lh,\quad \phi \mapsto \int_{\sem[\phi]_n} f_n^A. 
\]
Summing up, we have the composite map
\begin{equation}\label{eq:comparison-with-logic-on-words}\textstyle
\Fin(\sigma)\to \Mea(\P(\Typ_n(\sigma)))\to \Mea(\FO_n(\sigma)), \quad A\mapsto \int f_n^A\mapsto \mu_n^A.
\end{equation}
Essentially the same construction is featured in logic on words, \cf \ equation \eqref{def:R}:
\begin{itemize}[leftmargin=12pt]
\item The set of finite $\sigma$-structures $\Fin(\sigma)$ corresponds to the set of finite words $A^*$.
\item The collection $\Typ_n(\sigma)$ of (equivalence classes of) $\sigma$-structures with interpretations corresponds to $(A\times 2)^*$ or, interchangeably, $\beta(A\times 2)^*$ (in the case of one free variable).
\item The fragment $\FO_n(\sigma)$ of first-order logic corresponds to the Boolean algebra of languages, defined by formulas with a free variable, dual to the Boolean space $X$ appearing in \eqref{def:R}.
\item The first map in the composite \eqref{eq:comparison-with-logic-on-words} sends a finite $\sigma$-structure $A$ to the measure $\int f_n^A$ which, evaluated on $K\sue \Typ_n(\sigma)$, counts the (proportion of) interpretations $\alpha\colon \{v_1,\dots,v_n\}\to A$ such that $(A,\alpha)\in K$, similarly to $R$ from~\eqref{def:R}.
\item Finally, the second map in \eqref{eq:comparison-with-logic-on-words} sends a measure in $\Mea(\P(\Typ_n(\sigma)))$ to its pushforward along $\sem_n\colon \FO_n(\sigma)\mono \P(\Typ_n(\sigma))$. This is the second map in the composition \eqref{def:R}.
\end{itemize}

\medskip
On the other hand, the assignment $A\mapsto \mu_n^A$ defined in \eqref{eq:comparison-with-logic-on-words} is also closely related to the classical Stone pairing.
Indeed, for every formula $\phi$ in $\FO_n(\sigma)$,
    \begin{align}
        \mu_n^A(\phi)
            = \sum_{E\in \sem[\phi]_n}{f_n^A(E)}
            = \sum_{E\in \sem[\phi]_n} \sum_{(A, \alpha) \in E}\left(\frac{1}{|A|^n}\right)\cc \nonumber \\
            = \left( \frac{|\{ \overline a\in A^n \mid A\models \phi(\overline a)|}{|A|^n}\right)\cc = (\SPc{\phi, A})\cc,\label{eq:Stone-pairing-F}
    \end{align}
 where $\SPc{\ARG,\ARG}$ is the classical $\ui$-valued Stone pairing (\cf \ the beginning of Section~\ref{s:stone-pairing}).   
In this sense, $\mu_n^A$ can be regarded as a $\Lh$-valued Stone pairing, relative to the fragment $\FO_n(\sigma)$.
Next, we show how to extend this to the full first-order logic $\FO(\sigma)$. First, we observe that---as in the classical case---the construction is invariant under extensions of the set of free variables:
\begin{lem}\label{l:meas-restr}
Given $m, n\in \N$ and $A\in \Fin(\sigma)$, if $m\geq n$ then $(\mu_{m}^A)_{\restriction\FO_n(\sigma)}=\mu_n^A$.
\end{lem}
\begin{proof}
 Every variable assignment $\alpha\colon \{v_1,\dots,v_n\}\to A$ has $|A|^{m-n}$ possible extensions $\alpha'\colon \{v_1,\dots,v_m\}\to A$, to account for the $m-n$ unused variables. Moreover, for every $\phi \in \FO_n(\sigma)$, $(A,\alpha) \models \phi$ if, and only if, $(A,\alpha') \models \phi$.
    Because $\mu_m^A$ and $\mu_n^A$ take only values of ${}\cc$-type, and $p\cc + q\cc = (p + q)\cc$, we see that
    \begin{align*}
        \mu_n^A(\phi)
            &= \sum \left\{ \left(\frac{1}{|A|^{n}}\right)\cc \mid  \alpha\colon\{v_1,\dots,v_n\}\to A \ete{and} (A,\alpha) \models \phi \right\} \\
            &= \sum \left\{ \left(\frac{|A|^{m-n}}{|A|^{m}}\right)\cc \mid  \alpha\colon \{v_1,\dots,v_n\}\to A \ete{and} (A,\alpha) \models \phi \right\} \\
            &= \sum \left\{ \left(\frac{1}{|A|^{m}}\right)\cc \mid  \alpha'\colon \{v_1,\dots,v_m\}\to A \ete{and} (A,\alpha') \models \phi \right\} = \mu_m^A(\phi).
            \qedhere
    \end{align*}
\end{proof}
The Lindenbaum--Tarski algebra of all first-order formulas $\FO(\sigma)$ is the directed colimit of the chain of its Boolean subalgebras
\[
\FO_0(\sigma)\hookrightarrow \FO_1(\sigma)\hookrightarrow \FO_2(\sigma)\hookrightarrow \cdots.
\] 
Since the functor $\Mea$ turns directed colimits into codirected limits (Proposition~\ref{p:functor-mea}), the Priestley space $\Mea(\FO(\sigma))$ is the limit of the diagram
\[ \left\{
    \begin{tikzcd}[cramped, sep=2.5em]
        \Mea(\FO_n(\sigma)) & {\Mea(\FO_m(\sigma)) \mid m,n\in\N, \ m\geq n} \ar[swap, two heads]{l}{q_{n,m}}
    \end{tikzcd}
   \right\}
\]
where, for every measure $\mu\colon \FO_m(\sigma) \to \Lh$, the measure $q_{n,m}(\mu)$ is the restriction of $\mu$ to $\FO_n(\sigma)$. In view of Lemma~\ref{l:meas-restr}, for every $A\in\Fin(\sigma)$, the tuple $(\mu_n^A)_{n\in\N}$ is compatible with the restriction maps. Thus, recalling that limits in the category of Priestley spaces are computed as in sets, by universality of the limit construction this tuple yields a measure
\[ \SP{\ARG, A} \colon \FO(\sigma) \to \Lh \]
in the space $\Mea(\FO(\sigma))$ whose restriction to $\FO_n(\sigma)$ coincides with $\mu_n^A$, for every $n\in\N$. This we call the \emph{$\Lh$-valued Stone pairing} associated with $A$.
As in the classical case, using the fact that any two non-isomorphic finite structures are separated by a first-order sentence, it is not difficult to see that the mapping $A \mapsto \SP{\ARG, A}$ gives an embedding 
\[
\SP{\ARG,\ARG}\colon \Fin(\sigma) \mono \Mea(\FO(\sigma)).
\]

The following theorem illustrates the relation between the classical Stone pairing, namely $\SPc{\ARG,\ARG}\colon \Fin(\sigma)\mono \Meac(\FO(\sigma))$, and the $\Lh$-valued one.
\begin{thm}\label{t:embedding-triangle}
    The following diagram commutes:
    \begin{center}
        \begin{tikzcd}[row sep=small]
            {} & \Mea(\FO(\sigma))\ar[bend left=25]{dd}{\LG} \\
            \Fin(\sigma) \ar{ru}{\SP{\ARG,\ARG}}\ar[swap]{rd}{\SPc{\ARG,\ARG}} & \\
            & \Meac(\FO(\sigma))\ar[bend left=25]{uu}{\LC}
        \end{tikzcd}
    \end{center}
\end{thm}
\begin{proof}
    Fix an arbitrary finite $\sigma$-structure $A\in\Fin(\sigma)$. Let $\phi$ be a formula in $\FO(\sigma)$ with free variables among $\{v_1, \dots, v_n\}$, for some $n\in \N$. By construction, $\SP{\phi,A}=\mu_n^A(\phi)$. Therefore, by equation \eqref{eq:Stone-pairing-F}, $\SP{\phi,A}=(\SPc{\phi, A})\cc$ and thus the statement follows at once.
\end{proof}
\begin{rem}
The construction in this section works, mutatis mutandis, also for smaller logic fragments, \ie \ for sublattices $D\subseteq \FO(\sigma)$. 
This corresponds to composing the embedding $\Fin(\sigma)\mono\Mea(\FO(\sigma))$ with the restriction map $\Mea(\FO(\sigma)) \to \Mea(D)$ sending a measure $\mu\colon \FO(\sigma)\to\Lh$ to $\mu_{\restriction D}\colon D\to\Lh$. The only difference is that, in general, the ensuing map $\Fin(\sigma)\to \Mea(D)$ need not be injective.
\end{rem}

\subsection{Limits in the spaces of measures}\label{s:limits-finite-structures}
By Theorem \ref{t:embedding-triangle}, the $\Lh$-valued Stone pairing $\SP{\ARG,\ARG}$ and the classical Stone pairing $\SPc{\ARG,\ARG}$ determine each other. However, the notions of convergence associated with the spaces $\Mea(\FO(\sigma))$ and $\Meac(\FO(\sigma))$ are different: since the topology of $\Mea(\FO(\sigma))$ is richer, there are ``fewer'' convergent sequences.
Recall from Lemma~\ref{l:gamma-continuous-monotone} that $\LG\colon \Mea(\FO(\sigma)) \to \Meac(\FO(\sigma))$ is continuous. Also, $\LG (\SP{\ARG, A}) = \SPc{\ARG, A}$ by Theorem~\ref{t:embedding-triangle}. Thus, for any sequence of finite $\sigma$-structures $(A_n)_{n\in\N}$, if
\begin{center}
    $\SP{\ARG, A_n}$ converges to a measure $\mu$ in $\Mea(\FO(\sigma))$
\end{center}
then
\begin{center}
    $\SPc{\ARG, A_n}$ converges to the measure $\LG(\mu)$ in $\Meac(\FO(\sigma))$.
\end{center}

The converse is not true. For example, consider the signature $\sigma=\{<\}$ consisting of a single binary relation symbol, and let $(A_n)_{n\in\N}$ be the sequence of finite posets displayed in the picture below.
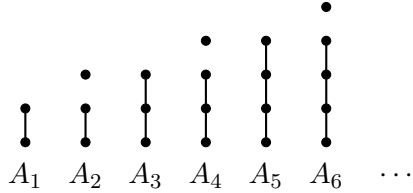
\begin{center}
\begin{tikzpicture}[xscale=0.80,yscale=0.45]
\tikzset{mynode1/.style={draw,circle,fill=black, inner sep=1.2pt,outer sep=0pt}}
    \node [mynode1, label={[yshift=-0.8cm]$A_1$}] (a1) at (0,0) {};
    \node [mynode1] (a2) at (0,1) {};

    \node [mynode1, label={[yshift=-0.8cm]$A_2$}] (b1) at (1,0) {};
    \node [mynode1] (b2) at (1,1) {};
    \node [mynode1] (b3) at (1,2) {};

    \node [mynode1, label={[yshift=-0.8cm]$A_3$}] (c1) at (2,0) {};
    \node [mynode1] (c2) at (2,1) {};
    \node [mynode1] (c3) at (2,2) {};

    \node [mynode1, label={[yshift=-0.8cm]$A_4$}] (d1) at (3,0) {};
    \node [mynode1] (d2) at (3,1) {};
    \node [mynode1] (d3) at (3,2) {};
    \node [mynode1] (d4) at (3,3) {};

    \node [mynode1, label={[yshift=-0.8cm]$A_5$}] (e1) at (4,0) {};
    \node [mynode1] (e2) at (4,1) {};
    \node [mynode1] (e3) at (4,2) {};
    \node [mynode1] (e4) at (4,3) {};

    \node [mynode1, label={[yshift=-0.8cm]$A_6$}] (f1) at (5,0) {};
    \node [mynode1] (f2) at (5,1) {};
    \node [mynode1] (f3) at (5,2) {};
    \node [mynode1] (f4) at (5,3) {};
    \node [mynode1] (f5) at (5,4) {};

    \node [label={[yshift=-0.8cm]$\cdots$}] (g1) at (6.2,0) {};

    \draw [thick] (a1) -- (a2) (b1) -- (b2) (c1) -- (c2) -- (c3) 
    (d1) -- (d2) -- (d3) (e1) -- (e2) -- (e3) -- (e4)
    (f1) -- (f2) -- (f3) -- (f4); 
\end{tikzpicture}
\end{center}
Consider the formula 
\[
\psi(x)\coloneqq \forall y \, [\neg(x<y)] \wedge \exists z \, [\neg(z < x)\mee \neg(z = x)]
\] 
stating that $x$ is maximal but not the maximum in the order given by $<$.
Then, for the sublattice $D= \{\false, \psi,\true\}$ of $\FO(\sigma)$, it is not difficult to see that the sequences $\SP{\ARG, A_n}$ and $\SPc{\ARG, A_n}$ converge in $\Mea(D)$ and $\Meac(D)$, respectively. For instance, the sequence
\[
\left(\SP{\psi,A_1},\, \SP{\psi,A_2},\, \SP{\psi,A_3},\, \SP{\psi,A_4},\, \dots\right)
\]
yields the sequence
\[
\left(0\cc,(\tfrac{2}{3})\cc, 0\cc, (\tfrac{2}{4})\cc, 0\cc, (\tfrac{2}{5})\cc,\ldots\right),
\]
which converges to $0\cc$. Similarly, we see that $(\SPc{\psi,A_n})_{n\in \N}$ converges to $0$. However, if we consider the Boolean algebra $B=\{\false, \psi,\neg\psi,\true\}$, then the $\SPc{\ARG, A_n}$'s still converge whereas the $\SP{\ARG, A_n}$'s do not. Just observe that the following sequence does not converge in $\Lh$ because the odd terms converge to $1\cc$, while the even terms converge to $1\mm$:
\[
\left(\SP{\neg\psi, A_n})_{n\in\N}=(1\cc, (\tfrac{1}{3})\cc, 1\cc, (\tfrac{2}{4})\cc, 1\cc, (\tfrac{3}{5})\cc,\ldots\right).
\]
 Nevertheless, there is a sequence $\SP{\ARG, A'_n}$ which converges in $\Mea(B)$ to a measure whose image under $\LG$ coincides with the limit of the $\SPc{\ARG, A_n}$'s (\eg, take the subsequence $(A'_n)_{n\in\N}$ of $(A_n)_{n\in\N}$ consisting of the even terms). In the next theorem, we will see that this is a general fact. 

Identify $\Fin(\sigma)$ with a subset of $\Mea(\FO(\sigma))$ (resp.\ $\Meac(\FO(\sigma))$) through $\SP{\ARG,\ARG}$ (resp.\ $\SPc{\ARG,\ARG}$).
A central question in the theory of structural limits, \cf \ \cite{nevsetvril2016first}, is to determine the closure of $\Fin(\sigma)$ in $\Meac(\FO(\sigma))$, which consists precisely of the limits of sequences of finite structures.  The following theorem gives an answer to this question in terms of the corresponding question for $\Mea(\FO(\sigma))$.
\begin{thm}\label{t:closure-f-Fin}
  Let $\overline{\Fin(\sigma)}$ denote the closure of $\Fin(\sigma)$ in $\Mea(\FO(\sigma))$. Then the set $\LG(\overline{\Fin(\sigma)})$ coincides with the closure of $\Fin(\sigma)$ in $\Meac(\FO(\sigma))$.
\end{thm}
\begin{proof}
    Write $U$ for the image of $\SP{\ARG,\ARG}\colon \Fin(\sigma) \mono \Mea(\FO(\sigma))$, and $V$ for the image of $\SPc{\ARG,\ARG}\colon \Fin(\sigma) \mono \Meac(\FO(\sigma))$. 
    We must prove that $\LG(\overline{U})=\overline{V}$. By Theorem \ref{t:embedding-triangle}, we have $\LG(U)=V$. Recall that $\LG\colon \Mea(\FO(\sigma))\to\Meac(\FO(\sigma))$ is continuous (Lemma \ref{l:gamma-continuous-monotone}), and the spaces $\Mea(\FO(\sigma))$ and $\Meac(\FO(\sigma))$ are compact Hausdorff (Proposition \ref{p:measures-Priestley} and Corollary \ref{c:class-meas-compact-pospace}). Since continuous maps between compact Hausdorff spaces are closed, we obtain $\LG(\overline{U})=\overline{\LG(U)}=\overline{V}$.
\end{proof}

\section{The logic of measures}\label{s:logic-of-measures}
In view of Proposition~\ref{p:measures-Priestley} we know that, for any distributive lattice $D$, the space $\Mea(D)$ of $\Lh$-valued measures on $D$ is a Priestley space, hence it has a dual distributive lattice $\PFG(D)$. In this section, we show that $\PFG(D)$ can be represented as the Lindenbaum--Tarski algebra for a propositional logic $\PL D$ obtained from $D$ by adding probabilistic quantifiers. 

The set of propositional variables of the logic $\PL D$ consists of the symbols $\PrG p a$, for every $a\in D$ and $p\in\Q\cap \ui$. The logical connectives that we shall consider are $\true,\false, \wedge,\vee$, and the top and bottom elements of $D$ are denoted $\true^*$ and $\false^*$, respectively\footnote{To avoid cumbersome notation, we use the same symbols for infima in $D$ and conjunctions in $\PL D$, namely $\wedge$, and for suprema in $D$ and disjunctions in $\PL D$, namely $\vee$. It will be clear from the context to which concept we are referring.}.
The logic $\PL D$ is specified by the following rules, for all $p,q,r\in \Q\cap \ui$ and $a,b\in D$ (along with the usual ones for the connectives $\true,\false, \wedge,\vee$):

\medskip
\quad
\begin{minipage}{0.91\textwidth}
    \begin{enumerate}[label=(L\arabic*)]
    \setlength\itemsep{0.3em}
        \item\label{L1} $\PrG q a \vdash \PrG p a$ whenever $p \leq q$
        \item\label{L2} $\PrG q a \vdash \PrG q b$ whenever $a\leq b$
        \item\label{L3} $\true \vdash \PrG 0 \false^*$, $\true \vdash \PrG q \true^*$, and $\PrG p \false^* \vdash \false$ whenever $p>0$
        \item\label{L4} $\PrG p a\,\wedge\, \PrG q b \vdash \PrG{p+q-r}(a\vee b) \,\vee\, \PrG{r}(a\wedge b)$ whenever $0\leq p+q-r\leq 1$
        \item\label{L5} $\PrG{p+q-r}(a\vee b)\,\wedge\,\PrG{r}(a\wedge b) \vdash \PrG p a \,\vee\, \PrG q b$ whenever $0\leq p+q-r\leq 1$
    \end{enumerate}
\end{minipage}
\vspace{1em}

The logic $\PL D$ can be interpreted in the space of measures $\Mea(D)$ by setting, for every $\mu\in\Mea(D)$,
\begin{equation}\label{eq:semantic-meas}
\mu\models \PrG p a \ \Leftrightarrow \ \mu(a)\geq p\cc.
\end{equation}
This satisfaction relation extends in the obvious way to the closure of the set of propositional variables under finite conjunctions and finite disjunctions. For any two formulas $\phi,\psi$ in the signature of $\PL D$, we write $\phi\models \psi$ provided that
\[
\forall \mu\in\Mea(D),\ \ \mu\models \phi \text{ implies }\mu\models \psi.
\]
The interpretation of $\PL D$ in $\Mea(D)$ given in \eqref{eq:semantic-meas} is sound:

\begin{lem}[Soundness]\label{l:soundness-L1-L5}
If $\phi\vdash \psi$ is any rule in \ref{L1}--\ref{L5}, then $\phi\models \psi$.
\end{lem}
\begin{proof}
\ref{L1}--\ref{L3} are easily verified, and we leave the details to the reader. We spell out the case of \ref{L4}; the proof of \ref{L5} is the same, mutatis mutandis.

Fix arbitrary $a,b\in D$ and $p,q,r\in \Q\cap \ui$ such that $0\leq p+q-r\leq 1$. Let $\mu\in\Mea(D)$ be any measure satisfying $\mu\models \PrG p a\,\wedge\, \PrG q b$, \ie \ $\mu(a)\geq p\cc$ and $\mu(b)\geq q\cc$. We must prove that $\mu\models \PrG{p+q-r}(a\vee b) \,\vee\, \PrG{r}(a\wedge b)$, that is either $\mu(a\vee b)\geq (p+q-r)\cc$ or $\mu(a\wedge b)\geq r\cc$. Using the fact that $\mu$ is a measure, we get
\[
p\cc\miss \mu(a\wedge b)\leq \mu(a)\miss \mu(a\wedge b) \leq \mu(a\vee b)\mip \mu(b)\leq \mu(a\vee b)\mip q\cc. 
 \]
 If $\mu$ does not satisfy $\mu(a\wedge b)\geq r\cc$, \ie \ $\mu(a\wedge b)\leq r\mm$, then
 \[
 p\cc\miss r\mm \leq \mu(a\vee b)\mip q\cc
 \]
 and therefore 
 \[
     (p+q-r)\cc=(p\cc\miss r\mm)+q\cc \leq (\mu(a\vee b)\mip q\cc) + q\cc \leq \mu(a\vee b).\qedhere
 \]
\end{proof}

In Theorem~\ref{t:PFG(D)-dual-to-meas} below we will see that, in fact, $\PL D$ is complete with respect to this measure-theoretic semantics. In other words, $\PL D$ is the logic of the space of measures $\Mea(D)$. To start with, we let $\PFG(D)$ be the Lindenbaum--Tarski algebra of $\PL D$, that is the quotient of the free distributive lattice on the set 
\[
\left\{\PrG p a\mid p\in\Q\cap \ui, \ a\in D\right\}
\]
with respect to the congruence generated by the conditions \ref{L1}--\ref{L5}. In particular, recall from Section~\ref{s:type-spaces} that elements of $\PFG(D)$ are equivalence classes of formulas $\phi,\psi$ with respect to equiprovability in the logic $\PL D$. Moreover, considering representatives, we have $\phi\leq \psi$ in $\PFG(D)$ if, and only if, $\phi \vdash \psi$ holds in the logic $\PL D$.

In order to show that every prime filter of $\PFG(D)$ yields a measure in $\Mea(D)$, we need the following lemma:
\begin{lem}\label{lem:technical}
The following statements hold:
\begin{enumerate}
\item for every $x,y\in \Lh$ such that $y\leq x$, $x \miss y = \bigvee{\left\{ p\cc \mip q\cc \mid y < q\cc \leq p\cc \leq x\right\}}$;
\item for every $x\in \Lh$, the map $x\mip \blank\colon [0\cc,x]\to\Lh$ sends suprema to infima.
\end{enumerate}
\end{lem}
\begin{proof}
(1) \ Recall that, for every $x,y\in \Lh$ such that $y\leq x$, 
\begin{equation*}
x \miss y = \bigvee{\left\{ x \mip q\cc \mid y < q\cc \leq x\right\}}.
\end{equation*}
So, it suffices to show that, whenever $q\cc \leq x$,
\begin{equation*}
x\mip q\cc= \bigvee{\left\{p\cc\mip q\cc \mid q\cc \leq p\cc \leq x\right\}}.
\end{equation*}
If $x = r\cc$ for some $r\in\Q\cap\ui$, the supremum belongs to the set  $\{ p\cc \mip q\cc \mid q\cc \leq p\cc \leq x \}$, and it is precisely $r\cc\mip q\cc$. Thus, suppose $x = r\mm$ for some $r\in (0,1]$. To show the non-trivial inequality, pick $z\in \Lh$ such that $p\cc \mip q\cc\leq z$ whenever $q\cc \leq p\cc \leq x$. Note that $\gamma(x)-q\leq\gamma(z)$, for otherwise we could find $p\in\Q\cap\ui$ satisfying $q\leq p<\gamma(x)$ and $\gamma(z)<p-q$, and so $z<p\cc \mip q\cc$. It follows that
\[
x\mip q\cc=(\gamma(x)-q)\mm\leq \gamma(z)\mm\leq z.
\]

\noindent(2) \ Fix $x\in \Lh$. We must prove that, for every subset $\{y_i\mid i\in I\}\subseteq [0\cc, x]$,
\[
x\mip \bigvee_{i\in I}{y_i}= \bigwedge_{i\in I}{(x\mip y_i)}.
\]
By Lemma \ref{l:properties-of-mip}, the map $x\mip\blank\colon [0\cc,x]\to \Lh$ is lower continuous and order reversing. Since the interval $[0\cc,x]$ is closed in $\Lh$, it is a compact ordered space with respect to the induced order and topology (in fact, it is a Priestley space). The supremum $\bigvee_{i\in I}{y_i}$ then coincides with the limit of $\{y_i\mid i\in I\}$, regarded as a net in $[0\cc,x]$ (\cf \ \cite[Proposition VI.1.3]{Compendium1980}). By lower continuity of $x\mip\blank$,  we have that $x\mip \bigvee_{i\in I}{y_i}$, regarded as an element of $\Lh{}^\op$, is less than or equal to the supremum (computed in $\Lh{}^\op$) of the $x\mip y_i$'s. That is, 
\[
x\mip \bigvee_{i\in I}{y_i}\geq \bigwedge_{i\in I}{(x\mip y_i)}
\]
in $\Lh$. The other inequality follows at once from the fact that $x\mip\blank$ is order reversing.
    \end{proof}

\begin{prop}\label{p:homo-to-measure}
    Let $\Pf\sue \PFG(D)$ be a prime filter. The assignment
    \[ 
    a\mapsto \bigvee{\left\{q\cc \mid \PrG q a \in \Pf\right\}} \quad \text{ defines a measure } \ \mu_{\Pf}\colon D \to \Lh.
    \]
\end{prop}
\begin{proof}
Condition \ref{L3} entails at once that $\mu_{\Pf}(\false^*)=0\cc$ and $\mu_{\Pf}(\true^*)=1\cc$. Further, suppose $a,b\in D$ and $a\leq b$. Then $\{q\mid \PrG q a\in \Pf\}\subseteq \{q\mid \PrG q b\in\Pf\}$ by \ref{L2}, and thus
\[
\mu_{\Pf}(a)= \bigvee{\left\{q\cc \mid \PrG q a\in \Pf\right\}}\leq \bigvee{\left\{q\cc \mid \PrG q b\in \Pf\right\}}=\mu_{\Pf}(b).
\]
That is, $\mu_{\Pf}$ is monotone. This shows that $\mu_{\Pf}$ satisfies the first two conditions defining $\Lh$-valued measures (\cf \  Definition~\ref{d:measure}). We prove the first half of the third condition, as the other half is proved in a similar fashion. We must show that, for every $a,b\in D$,  
\begin{align} \label{eq:mu-h-1}
    \mu_{\Pf}(a)\miss \mu_{\Pf}(a\wedge b)\leq \mu_{\Pf}(a\vee b)\mip \mu_{\Pf}(b).
\end{align}
    By item 1 in Lemma~\ref{lem:technical}, the left-hand expression in~\eqref{eq:mu-h-1} can be written as
    \[
    \mu_{\Pf}(a)\miss \mu_{\Pf}(a\wedge b) = \bigvee{\left\{p\cc\mip r\cc\mid \mu_{\Pf}(a\wedge b)<r\cc\leq p\cc\leq\mu_{\Pf}(a)\right\}}.
    \]
    Furthermore, we have $\mu_{\Pf}(b)=\bigvee{\left\{q\cc \mid q\cc \leq \mu_{\Pf}(b)\right\}}$. Just observe that $\PrG q b \in \Pf$ entails $q\cc \leq \mu_{\Pf}(b)$, and so
    \[
    \mu_{\Pf}(b)=\bigvee{\left\{q\cc \mid \PrG q b \in \Pf\right\}}\leq \bigvee{\left\{q\cc \mid q\cc \leq \mu_{\Pf}(b)\right\}}\leq \mu_{\Pf}(b).
    \]
    By item 2 in Lemma~\ref{lem:technical}, the right-hand expression in~\eqref{eq:mu-h-1} is equivalently expressed as
    \begin{align*}
    \mu_{\Pf}(a\vee b)\mip \mu_{\Pf}(b) &= \mu_{\Pf}(a\vee b)\mip \bigvee{\left\{q\cc \mid q\cc \leq \mu_{\Pf}(b)\right\}} \\
    &=\bigwedge{\left\{\mu_{\Pf}(a\vee b)\mip q\cc \mid q\cc \leq \mu_{\Pf}(b)\right\}}.
    \end{align*}

   Therefore, we must prove that
    \begin{equation*}
    \bigvee{\left\{p\cc\mip r\cc\mid \mu_{\Pf}(a\wedge b)<r\cc\leq p\cc\leq\mu_{\Pf}(a)\right\}}\leq \bigwedge{\left\{\mu_{\Pf}(a\vee b)\mip q\cc\mid q\cc \leq \mu_{\Pf}(b)\right\}}.
    \end{equation*}
That is, whenever $\mu_{\Pf}(a\wedge b)<r\cc\leq p\cc\leq \mu_{\Pf}(a)$ and $q\cc \leq \mu_{\Pf}(b)$, we have 
\[
(p-r)\cc \leq \mu_{\Pf}(a\vee b)\mip q\cc.
\]
In view of Lemma~\ref{lem:mip-is-right-adjoint}, the latter inequality is equivalent to $(p+q-r)\cc\leq \mu_{\Pf}(a\vee b)$. 
In turn, using \ref{L4} and the fact that $\Pf$ is a prime filter, $\PrG p a,\PrG q b\in \Pf$ and $\PrG r (a\wedge b)\notin \Pf$ entail $\PrG{p+q-r}(a\vee b)\in \Pf$. Hence,
\[
\mu_{\Pf}(a\vee b)=\bigvee{\left\{s\cc \mid \PrG s (a\vee b)\in \Pf\right\}}\geq (p+q-r)\cc. \qedhere
\]
\end{proof}

We are now in a position to show that the dual lattice of the Priestley space $\Mea(D)$ is isomorphic to the Lindenbaum--Tarski algebra $\PFG(D)$ for the logic $\PL D$, built from the propositional variables $\PrG p a$ by imposing the laws \ref{L1}--\ref{L5}. This amounts to the completeness of $\PL D$ with respect to its interpretation in $\Mea(D)$. 
\begin{thm}[Completeness]\label{t:PFG(D)-dual-to-meas}
   Let $D$ be a distributive lattice. Then the lattice $\PFG(D)$ is isomorphic to the distributive lattice dual to the Priestley space $\Mea(D)$.
\end{thm}
\begin{proof}
Let $X_{\PFG(D)}$ be the Priestley space dual to $\PFG(D)$. By Proposition \ref{p:homo-to-measure}, there is a map $\theta\colon X_{\PFG(D)}\to \Mea(D)$, $\Pf\mapsto \mu_{\Pf}$. We claim that $\theta$ is an isomorphism of Priestley space, \ie \ a continuous bijection that is also an order embedding. The statement then follows by duality.

It is not difficult to see that $\theta$ is monotone. Just observe that $F_1\subseteq F_2$ implies 
\[
\mu_{\Pf_1}(a)=\bigvee{\left\{q\cc \mid \PrG q a \in \Pf_1\right\}}\leq \bigvee{\left\{q\cc \mid \PrG q a \in \Pf_2\right\}}=\mu_{\Pf_2}(a)
\]
for all $a\in D$. Moreover, if $\mu_{\Pf_1}(a)\not\leq \mu_{\Pf_2}(a)$ for some $a\in D$, then
   \begin{equation}\label{eq:inequality-embed}
   \bigvee{\left\{q\cc\mid \PrG q a\in \Pf_1\right\}}=\mu_{\Pf_1}(a)\not\leq \mu_{\Pf_2}(a) = \bigwedge{\left\{p\mm\mid \PrG p a\notin \Pf_2\right\}},
   \end{equation}
   where the last equality follows from the following fact:
   \begin{clm}
   For every prime filter $F\in X_{\PFG(D)}$, $\bigvee{\left\{q\cc\mid \PrG q a\in \Pf\right\}}=\bigwedge{\left\{p\mm\mid \PrG p a\notin \Pf\right\}}$.
   \end{clm}
   \begin{proof}[Proof of Claim]
   Fix an arbitrary $a\in D$. By \ref{L1}, it follows at once that 
\[ 
  \bigvee{\left\{q\cc \mid \PrG q a \in \Pf\right\}}\leq  \bigmee{\left\{p\mm \mid \PrG p a\notin \Pf\right\}}.
    \]
    Suppose, by contradiction, that the inequality is strict. Then there must be an $r\in\Q\cap (0,1]$ such that 
    \[
    \bigvee{\left\{q\cc \mid \PrG q a \in \Pf\right\}}=r\mm<r\cc=\bigmee{\left\{p\mm \mid \PrG p a\notin \Pf\right\}}.
    \]
    Just observe that $\left\{q\mid \PrG q a \in \Pf\right\}$ and $\left\{p\mid \PrG p a\notin \Pf\right\}$ form a partition of $\Q\cap \ui$.
    If $\PrG r a \in \Pf$, then $r\cc \leq \bigvee{\left\{q\cc \mid \PrG q a\in\Pf\right\}}= r\mm$, a contradiction. Similarly, if $\PrG r a\notin \Pf$ we get $r\cc=\bigmee{\left\{p\mm \mid \PrG p a\notin\Pf\right\}}\leq r\mm$, a contradiction.
   \end{proof}
   Equation \eqref{eq:inequality-embed} implies the existence of $p,q$ satisfying $\PrG q a\in \Pf_1$, $\PrG p a\notin \Pf_2$ and $q\cc\not\leq p\mm$, \ie \ $q\geq p$. It follows by \ref{L1} that  $\PrG p a\in \Pf_1$. We conclude that $\PrG p a \in \Pf_1 \setminus \Pf_2$, hence $\Pf_1\not\sue \Pf_2$.  This shows that $\theta$ is an order embedding. 
   
   Next, we prove that $\theta$ is surjective. Fix a measure $\mu\in\Mea(D)$. It is not hard to see, using Lemma \ref{l:soundness-L1-L5}, that the filter $\Pf_{\mu}\sue \PFG(D)$ generated by
   \[
       \left\{ \PrG q a \mid a\in D,\ q\in\Q\cap\ui,\  q\cc\leq \mu(a)\right\}
   \]
   is prime. Further,  
    $\theta(\Pf_{\mu})(a)=\bigvee{\left\{q\cc\mid \PrG q a \in \Pf_\mu\right\}}=\bigvee{\left\{q\cc\mid q\cc\leq \mu(a)\right\}} =\mu(a)$ for every $a\in D$. Hence, $\theta(\Pf_{\mu})=\mu$ and $\theta$ is surjective. Since $\theta$ is an order embedding, it is also injective, and thus a bijection.
    
    To settle the theorem, it remains to show that $\theta$ is continuous.
For every subset of $\Mea(D)$ of the form $\MBU{a}{p}=\left\{\mu\in\Mea(D)\mid \mu(a)\geq p\cc\right\}$ where $a\in D$ and $p\in\Q\cap \ui$, 
\begin{align*}
\theta\inv(\MBU{a}{p})&=\left\{\Pf\in X_{\PFG(D)}\mid \mu_{\Pf}(a)\geq p\cc\right\} \\
&=\left\{\Pf\in X_{\PFG(D)}\mid \bigvee{\left\{q\cc\mid \PrG q a\in \Pf\right\}}\geq p\cc\right\} \\
&=\left\{\Pf\in X_{\PFG(D)}\mid \PrG p a \in \Pf\right\}
\end{align*}
    which is an open subset of $X_{\PFG(D)}$.
     Similarly, for a subset of $\Mea(D)$ of the form $\MBD{a}{q} = \left\{\mu\in\Mea(D)\mid \mu(a)\leq q\mm\right\}$ with $a\in D$ and $q\in\Q\cap (0,1]$, in view of the Claim above we have that $\theta\inv(\MBD{a}{q})=\left\{\Pf\in X_{\PFG(D)}\mid \PrG q a\notin \Pf\right\}$, which is again open in $X_{\PFG(D)}$. Hence, $\theta$ is continuous.
\end{proof}

\paragraph*{\bf Adding Boolean complements.}
By Theorem~\ref{t:PFG(D)-dual-to-meas}, for any distributive lattice $D$, the lattice of clopen up-sets of $\Mea(D)$ is isomorphic to the Lindenbaum--Tarski algebra $\PFG(D)$ of our ``positive'' propositional logic $\PL D$. Moving from the lattice of clopen up-sets to the Boolean algebra of \emph{all} clopens logically corresponds to adding negation to the logic.
The logic obtained in this way can be presented as follows.

Introduce a new propositional variable $\PrL q a$ for each $a\in D$ and $q\in \Q\cap\ui$. We add, along with the propositional connective $\neg$, also a new rule to the logic $\PL D$, stating that $\PrL q a$ is the negation of $\PrG q a$:

\smallskip
\quad
\begin{minipage}{0.8\textwidth}
    \begin{enumerate}[label=(L\arabic*)]\setcounter{enumi}{5}
        \item\label{L6} $\PrL q a \mee \PrG q a\vdash \false$ and $\true \vdash \PrL q a \vee \PrG q a$
    \end{enumerate}
\end{minipage}
\smallskip

\noindent (Equivalently, $\PrG q a\vdash \neg \PrL q a$ and $\neg \PrL q a\vdash \PrG q a$.) Write $\mathcal{PL}^*_D$ for the ensuing logic.

For a measure $\mu\in\Mea(D)$, the new propositional variables are interpreted as
\[
\mu\models \PrL q a \ \Leftrightarrow \ \mu(a)< q\cc.
\]

Clearly, \ref{L6} is satisfied in $\Mea(D)$. Moreover, it is not difficult to see that the Boolean algebra of all clopen subsets of $\Mea(D)$ is isomorphic to the quotient of the free distributive lattice on
\[
   \left\{\PrG p a\mid p\in\Q\cap \ui, \ a\in D\} \cup \{\PrL q b\mid q\in\Q\cap \ui, \ b\in D\right\}
\]
with respect to the congruence generated by \ref{L1}--\ref{L6}. In turn, the latter is isomorphic to the Lindenbaum--Tarski algebra of the logic $\mathcal{PL}^*_D$. That is, $\mathcal{PL}^*_D$ is complete with respect to its interpretation in $\Mea(D)$.

\paragraph*{\bf Specialising to $\FO(\sigma)$.}
Let us briefly discuss what happens when we instantiate $D$ with the full first-order logic $\FO(\sigma)$. For a formula $\phi\in \FO(\sigma)$ with free variables among $v_1, \dots, v_n$ and a rational $q\in \Q\cap\ui$, we have two new sentences $\PrG q \phi$ and $\PrL q \phi$. For a finite $\sigma$-structure $A$ identified---via the $\Lh$-valued Stone pairing---with the measure $\SP{\ARG,A}$,
\[ 
A \models \PrG q \phi \qtq{if, and only if,} \SP{\phi, A} \geq q\cc\]
and
\[
A \models \PrL q \phi \qtq{if, and only if,} \SP{\phi, A} < q\cc.
\]
That is, $\PrG q \phi$ is true in $A$ precisely when a random assignment of the variables $v_1, \dots, v_n$ in $A$ satisfies $\phi$ with probability at least $q$, and similarly for $\PrL q \phi$. 

For example, in the signature of (directed) graphs, \ie \ when $\sigma$ consists of one binary relation $R(\ARG,\ARG)$, the formula $\PrG{\frac{1}{2}} R(x,x)$ in $\PL{\FO(\sigma)}$ expresses the property that at least half of the vertices have a loop.

The following is a special instance of the results presented above:

\begin{thm}\label{t:axiomatisation-Boolean-case}
The Boolean space $\Mea(\FO(\sigma))$ is dual to the Lindenbaum--Tarski algebra of the propositional logic having as atoms $\PrG q \phi$ and $\PrL q \phi$, for each $\phi \in \FO(\sigma)$ and $q\in \Q\cap\ui$, and the following inference rules (along with the usual ones for the Boolean connectives):
\begin{equation*}
\begin{gathered}
\infer[{\scriptstyle(\mathrm{if} \ p\, \leq \, q)}]
{\PrG p \phi}{\PrG q \phi}
\hspace{2em}
\infer[{\scriptstyle(\mathrm{if} \ \phi \, \vdash \, \psi)}]
{\PrG q \psi}{\PrG q \phi}
\hspace{2em}
\infer
{\PrG 0 \bot}{}
\hspace{2em}
\infer[{\scriptstyle(\mathrm{if} \ q \, > \, 0)}]
{\PrL q \bot}{}
\hspace{2em}
\infer
{\PrG q \top}{}
\hspace{2em}
\infer=
{\neg \PrL q \phi}{\PrG q \phi} \\[2ex]
\infer
{\PrG{p+q-r}(\phi\vee \psi) \,\vee\, \PrG{r}(\phi\wedge \psi)}{\PrG p \phi\,\wedge\, \PrG q \psi}
\hspace{2em}
\infer[{\hspace{0.4em}\scriptstyle(\mathrm{if} \ 0 \, \leq \, p+q-r \, \leq \, 1)}]
{\PrG p \phi \,\vee\, \PrG q \psi}{\PrG{p+q-r}(\phi\vee \psi)\,\wedge\,\PrG{r}(\phi\wedge \psi)}
\end{gathered}
\end{equation*}
\end{thm}

\medskip

Therefore, if we regard $\PrG q$ and $\PrL q$ as probabilistic quantifiers that bind all free variables of a given formula, then the Stone pairing $\SP{\ARG,\ARG}\colon \Fin(\sigma) \to \Mea(\FO(\sigma))$ can be seen as the embedding of finite structures into the space of $0$-types for the logic with probabilistic quantifiers axiomatised in the previous theorem.


\section{Related work}
Our logic with probabilistic quantifiers (\cf \ Theorem~\ref{t:axiomatisation-Boolean-case}) is related to other axiomatisations of logics arising from space-of-measures constructions which have appeared in the literature. At the time of submitting this paper the authors were not aware of the almost identical axiomatisation of the dual of the probabilistic powerspace construction in the setting of stably compact spaces, due to Moshier and Jung~\cite{moshierjung2002logic}. Their work is heavily inspired by the axiomatisation given by Heckmann~\cite{heckmann1993probabilistic} in terms of a certain geometric propositional logic. The work of Heckmann was later streamlined by Vickers~\cite{Vickers2008}. For another infinitary duality theoretic approach to logics with probabilistic quantifiers see \eg \ \cite{KLMP2013,FKLMP2017}, where an axiomatisation based on the signature of $\sigma$-complete Boolean algebras is provided.

Logics with probabilistic quantifiers have been studied also in the setting of finite model theory. In fact, Friedman's probabilistic quantifiers (cf.~\cite{steinhorn1985chapter,steinhorn1985borel}), which express that the probability is $>0$, are the main ingredient in the proof of one of the main results in the theory of structural limits, see \cite[Theorem~3.2 and Corollary~4.3]{nevsetvrilmendez2019existence}. Furthermore, Kontinen~\cite{kontinen2010coherence}, Keisler and Lotfallah~\cite{keisler2009almost}, and Knyazev~\cite{knyazev1990zero} showed that probabilistic quantifiers behave reasonably well with respect to zero-one laws.

Probabilistic quantifiers akin to ours were extensively studied already in the 1970s, see Keisler's survey~\cite{keisler1985chapter}. The main difference, compared to our approach, is that infinitary conjunctions or rules need to be added in order to obtain completeness, and therefore the compactness theorem is lost. Our logic, on the other hand, does not allow for nesting of probabilistic quantifiers, is finitary and satisfies the compactness theorem. The latter follows at once from duality theory and the fact that the space of $\Lh$-valued measures dual to $\mathcal{PL}^*_D$ is topologically compact. More recently, Zhou~\cite{zhou2009complete} showed that a propositional version of Keisler logic can be made finitary; however, this logic still fails to satisfy the compactness theorem.

\section{Conclusion}
Types are points of the dual space of a logic (identified with its Lindenbaum--Tarski algebra). In classical first-order logic, $0$-types are just the models modulo elementary equivalence. But when there are not ``enough'' models, as in finite model theory, the spaces of types provide completions of the sets of models.

In \cite{GPR2017}, it was shown that for logic on words and various quantifiers we have that, given a Boolean algebra of formulas with a free variable, the space of types of the Boolean algebra generated by the formulas obtained by quantification is given by a space-of-measures construction. Here we have shown that a suitable enrichment of first-order logic gives rise to a space of measures $\Mea(\FO(\sigma))$ closely related to the space $\Meac(\FO(\sigma))$ used in the theory of structural limits. Indeed, Theorem~\ref{t:embedding-triangle} tells us that the ensuing Stone pairings interdetermine each other. Further, the Stone pairing for $\Mea(\FO(\sigma))$  is just the embedding of the finite models in the completion/compactification provided by the space of $0$-types of the enriched logic (Theorem~\ref{t:axiomatisation-Boolean-case}).

These results identify the logical gist of the theory of structural limits, and provide a new and interesting connection between logic on words and the theory of structural limits in finite model theory. But we also expect that it may prove a useful tool in its own right. 

For instance, for structural limits, it is an open problem to characterise the closure $\overline{\Fin(\sigma)}$ of the image of the $\ui$-valued Stone pairing \cite{nevsetvril2016first}. The same question translates equivalently to the $\Lh$-valued setting (Theorem~\ref{t:closure-f-Fin}), native to logic and where we can use duality.
The closure $\overline{\Fin(\sigma)}$, seen as a subspace of $\Mea(\FO(\sigma))$, is also a Priestley space and this embedding dually corresponds to a quotient of the Lindenbaum--Tarski algebra of the enriched logic.
Understanding this quotient is crucial in order to characterise $\overline{\Fin(\sigma)}$. One would expect that this is the subspace $\Mea(\FO(T_\text{\em fin}))$ of $\Mea(\FO(\sigma))$ given by the quotient $\FO(\sigma)\twoheadrightarrow \FO(T_\text{\em fin})$ onto the theory of pseudofinite structures. The purpose of such a characterisation would be to understand the points of the closure as ``generalised models". 

Another subject that we would like to investigate is that of zero-one laws. The zero-one law for first-order logic states that the sequence of measures for which the $n$th measure, on a sentence $\psi$, yields the proportion of $n$-element structures satisfying $\psi$, converges to a $\{0,1\}$-valued measure. Over $\Lh$ this will no longer be true as $1$ is split into its ``limiting'' and ``achieved'' personae.  
Yet, we expect the above sequence to converge also in this setting and, by Theorem~\ref{t:embedding-triangle}, it will converge to a $\{0\cc,1\mm,1\cc\}$-valued measure. 
Understanding this finer grained measure may yield useful information about the zero-one law.

Finally, it would be interesting to investigate whether the limits for schema mappings introduced by Kolaitis \emph{et al.} \cite{Kolaitis2018} may be seen also as a type-theoretic construction. Also, we would want to explore the connections with other semantically inspired approaches to finite model theory, such as those recently put forward by Abramsky, Dawar \emph{et al.} \cite{Abramsky2017b,AbramskyShah2018}.

\paragraph{Acknowledgements}
The authors are grateful to the anonymous referees for their valuable comments and suggestions.

\bibliographystyle{alphaurl}

\appendix

\section{Extended Priestley duality and canonical extensions}\label{s:ext-prie-dual-can-ext}

The purpose of this section is to provide the necessary background on extended Priestley duality so that, in Appendix~\ref{s:apply-ext-prie-dual}, we can justify the definition of $\Lh$ and its structure. There is a rich literature on extensions of the duality between distributive lattices and Priestley spaces to dualities for lattices with additional operations, see \eg \ \cite{Goldblatt1989}. The prototypical example (in the case of Boolean algebras) is the celebrated J\'onsson--Tarski duality \cite{JT1951,JT1952}, from which most of the theory originates.

We present an algebraic approach to extended Priestley duality which is based on the theory of canonical extensions. Its main feature is that both the space $X$ and its dual lattice $A$ embed (order-theoretically) into a complete lattice $A\can$, the \emph{canonical extension} of $A$. This allows us to capture Priestley duality in purely algebraic terms, and is an incredibly powerful approach for dealing with additional operations on lattices. For example, if $A$ comes equipped with a binary operation $h\colon A\times A\to A$ then, by studying the relationship between the graph of $h$ and the points of $X$ in the product $A\can \times A\can \times A\can$, we can identify two partial binary operations on $X$ that fully capture $h$. An instance of this situation is studied in this section. Our main inspiration comes from \cite{GP2007,GP2007b} and also \cite{FGGM20}.

\subsection{Basic notation and terminology}
The canonical extension of a (bounded) distributive lattice $A$ is an embedding $e\colon A \mono A\can$ of $A$ into a complete lattice $A\can$ that is \emph{dense} and \emph{compact}:
\begin{enumerate}[labelindent=4em, leftmargin=*, itemsep=0.8em, topsep=1em]
    \item[(Dense)] Every element of $A\can$ is both a join of meets and a meet of joins of elements in the image of $e$.
    \item[(Compact)] Given subsets $S$ and $T$ of $A$ with $\bigwedge{e(S)}\leq \bigvee{e(T)}$, there are finite sets $S'\subseteq S$ and $T'\subseteq T$ such that $\bigwedge{e(S')}\leq \bigvee{e(T')}$.
\end{enumerate}
Canonical extensions of distributive lattices always exist and are unique, up to isomorphism. In fact, $A\can$ is isomorphic to the lattice of all up-sets of the Priestley space dual\footnote{Note that, with respect to the convention of Remark~\ref{r:order-issues} below, $A\can$ is actually isomorphic to the down-sets of the dual space.} to $A$. In the following, we will always identify $A$ with a sublattice of its canonical extension $A\can$.

\emph{Open elements} of $A\can$ are those of the form $\bigvee S$, for some $S\sue A$. The set of all open elements of $A\can$ is denoted $\I(A\can)$. In fact, there is a bijection between open elements of $A\can$ and ideals of $A$, given by $i\in \I(A\can) \mapsto \downset i \cap A$ and $I\sue A \mapsto \bigvee I$. The latter restricts to a bijection between the set $\M(A\can)$ of completely meet-irreducible elements of $A\can$ and prime ideals of $A$. Recall that $p\in A\can$ is \emph{completely meet-irreducible} if, for all $S \subseteq C$, $\bigmee S \leq p$ implies $a \leq p$ for some $a\in S$. The set $\F(A\can)$ of \emph{closed elements} of $A\can$, and the set $\J(A\can)$ of \emph{completely join-irreducible elements} of $A\can$ are defined order-dually.

Note that $\M(A\can)$ does not include the top element of $A\can$. However, it is sometimes useful to add it, and so we will denote by $\M_1(A\can)$ the set $\M(A\can)\cup \{1\}$ and, similarly, $\J_0(A\can)$ denotes the set $\J(A\can)\cup \{0\}$.

For a distributive lattice $A$, every element of $A\can$ is a meet of completely meet-irreducible elements and a join of completely join-irreducible elements. Moreover, if we equip $\J(A\can)$ and $\M(A\can)$ with the partial orders induced by $A\can$, we have an order-isomorphism
\[ \kappa\colon \J(A\can) \to \M(A\can), \ \ j\mapsto \bigvee{\{a\in A\mid j\not\leq a\}} \]
satisfying, for every $m\in \M(A\can)$, $j\in \J(A\can)$ and $u\in A\can$,
\[ u \leq \kappa(j) \iff j\nleq u \qtq{and} \ki(m) \leq u \iff u\nleq m. \]
This corresponds to the fact that the complement of a prime filter is a prime ideal, and vice versa. For more details, we refer the reader to~\cite{GJ2004}.

\subsection{Priestley duality}\label{s:priestley-duality}
Thanks to the correspondence between prime filters (resp.\ prime ideals) of a distributive lattice $A$ and completely join-irreducible (resp.\ completely meet-irreducible) elements of $A\can$, we can retrieve the Priestley space $X = (X,\leq,\tau)$ dual to $A$ by endowing either the set $\M(A\can)$, or the set $\J(A\can)$, with an order and topology. Both choices result in the same space, but we opt for the former as it provides a minor technical advantage in our applications. The order on $\M(A\can)$ is inherited from $A\can$ and the subbase of the topology is formed by the sets $\{ m \mid a \nleq m \}$ and their complements, for every $a\in A$.

\begin{rem}\label{r:order-issues}
We warn the reader that, with the definition of dual Priestley space given in Section~\ref{s:duality}, the Priestley space $X$ just defined is, in fact, dual to the lattice $A^\op$ and not $A$. Just observe that, for $m,n\in \M(A\can)$,
\[ m \leq n \ee\iff I_m \sue I_n  \ee\iff F_{\ki(m)} \supseteq F_{\ki(n)}, \]
where $I_m \coloneqq \{ a \in A \mid a \leq m \}$ is the prime ideal corresponding to $m\in \M(A\can)$ and $F_j \coloneqq \{ a \in A \mid j \leq a \}$ is the prime filter corresponding to $j\in \J(A\can)$. When working with canonical extensions, it is convenient to order prime filters by \emph{reverse} inclusion, so that both $X$ and its dual lattice $A$ order-theoretically embed into $A\can$. However, we opted for the inclusion order in the main text because it gives the usual pointwise order between measures. To apply the general theory to $\Lh$, in Appendix~\ref{s:apply-ext-prie-dual} we will take order-duals on the lattice side.
\end{rem}

To also capture the duality between lattice homomorphisms and Priestley morphisms, we recall how maps between lattices extend to maps between their canonical extensions. For a monotone map $f\colon A\to B$ between distributive lattice, its \emph{sigma extension} $f^\sigma\colon A\can \to B\can$ is defined by
\begin{align*}
    f^\sigma(k) &= \bigmee \{ f(a) \mid a\in A \ete{and} k \leq a\} \quad\text{for all } k\in \F(A\can)\\
    f^\sigma(u) &= \bigvee \{ f^\sigma(k) \mid k\in \F(A\can) \ete{and} k \leq u\} \quad\text{for all } u\in A\can
\end{align*}
and, dually, its \emph{pi extension} $f^\pi\colon A\can \to B\can$ is defined by
\begin{align*}
    f^\pi(i) &= \bigvee \{ f(a) \mid a\in A \ete{and} a \leq i\} \quad\text{for all } i\in \I(A\can)\\
    f^\pi(u) &= \bigmee \{ f^\pi(i) \mid i\in \I(A\can) \ete{and} u \leq i\} \quad\text{for all } u\in A\can.
\end{align*}
The two extensions are closely related as they agree on open and closed elements. Moreover, if $f$ preserves binary joins then $f^\sigma$ preserves all non-empty joins and, similarly, if $f$ preserves binary meets then $f^\pi$ preserves all non-empty meets.

Whenever $f^\sigma = f^\pi$, we say that $f$ is \emph{smooth} and denote the unique extension by $f\can$. This happens, for example, when $f$ preserves finite (possibly empty) joins or finite meets.
Recall that a monotone map $f$ between complete lattices preserves all joins if and only if it has a right adjoint $f\ra$ and, dually, it preserves all meets precisely when it has a left adjoint $f\la$. Moreover, if $h\colon A\to B$ is a lattice homomorphism, the right adjoint $h\can{}\ra$ of $h\can\colon A\can \to B\can$ preserves completely meet-irreducible elements and, dually, the left adjoint  $h\can{}\la$ preserves completely join-irreducible elements. We can compute the continuous monotone map dual to a homomorphism $h\colon A\to B$ as the restriction of $h\can{}\ra \colon B\can \to A\can$ to the completely meet-irreducible elements. Note that, if we defined dual Priestley spaces as based on $\J(A\can)$, we would have to take the restriction of $h\can{}\la \colon B\can \to A\can$ to completely join-irreducible elements.

\subsection{Duality theory for operators}

In this section we outline some facts about the duality theory of normal double quasioperators $A^n \to A$ based on \cite{GP2007,GP2007b}. Double quasioperators are double operators where arguments can have different monotonicity types. In order to avoid the notational difficulties that arise when presenting the theory for different monotonicity types, we assume that we have a three-sorted binary operator that is monotone in both coordinates.

\begin{defi}
    Let $A, B$ and $C$ be distributive lattices.
    We say that a map $h\colon A\times B \to C$ is a \emph{three-sorted operator} if
\begin{itemize}
    \item $h$ is a double operator, \ie \ $h$ preserves binary meets and binary joins in both coordinates,
    \item $h$ is dual-normal, \ie \ $h(a,1) = 1 = h(1,b)$ for every $a\in A$ and $b\in B$, and
    \item $h(0,0) = 0$.
\end{itemize}
\end{defi}

\begin{rem}
    Everything true of three-sorted operators can also be applied to the dual scenario as well, that is, to the double operators $g\colon A\times B\to C$ such that $g(a,0) = 0 = g(0,b)$ and $g(1,1) = 1$. In this case, we simply treat $g$ as a three-sorted operator of the type $A^\op \times B^\op \to C^\op$. Likewise, the same definition can be used to deal with different monotonicity types, such as a double operator $A\times B^\op\to C$ that is antitone in the second coordinate. Here, only the coordinates that are antitone have to be considered dually.
\end{rem}

Throughout this section we fix a three-sorted operator $h\colon A\times B \to C$. We denote the Priestley spaces dual to the distributive lattices $A$, $B$ and $C$ by $X$, $Y$ and $Z$, respectively. Because $h\colon A\times B \to C$ preserves finite meets (including the empty meet $1$) and binary joins in both coordinates, its pi extension $f \coloneqq h^\pi\colon A\can \times B\can \to C\can$ preserves all meets in both coordinates and its sigma extension $g \coloneqq h^\sigma\colon A\can \times B\can \to C\can$ is a complete operator (\ie \ it preserves all non-empty joins in both coordinates).

We will also make use of the map $\zer\colon A \to C$ defined as $a\mapsto h(a,0)$. Observe that it preserves finite meets and finite joins, hence it is a lattice homomorphism. The continuous monotone map $i\colon Z \to X$ dual to $\zer\colon A \to C$ is simply the restriction of the right adjoint of $\zer\can\colon A\can \to C\can$ to $\M(C\can) \to \M(A\can)$.

\begin{lem}\label{l:amp-f-g}
    For every $u\in \I(A\can)\cup \F(A\can)$ we have $\zer\can u = f(u,0) = g(u,0)$.
\end{lem}
\begin{proof}
    For $i\in \I(A\can)$, $f(i,0) = \bigvee \{ h(a,0) \mid a\in A \ete{and} a\leq i \} = \zer^\pi i = \zer\can i$ because $(A \times B)\can \cong A\can \times B\can$ and $\I(A\can \times B\can) \cong \I(A\can)\times \I(B\can)$. Also, $\zer\can i = g(i,0)$ holds because $f$ and $g$ agree on open and closed elements. The same proof, mutatis mutandis, shows that $\zer\can k = g(k,0) = f(k,0)$ for all $k\in \F(A\can)$.
\end{proof}

\subsection{The first partial operation}\label{s:duality-for-mip}
Traditionally, given a distributive lattice with a quasioperator, one endows the dual Priestley space with two ternary relations satisfying certain topological and order-theoretic properties which uniquely characterise the quasioperator. As was already evident in \cite{GP2007,GP2007b} (and successfully applied in \cite{FGGM20}), these two relations arise from two partial operations, by taking the upwards or downwards closures of the codomains.

The first partial operation represents the pi extension of our operator $h\colon A\times B\to C$. Fix $u\in A\can$. Since $f(u,\ARG)$ preserves all meets, it has a left adjoint $f\la_1\colon A\can \times C\can \to B\can$, \ie
\[ f\la_1(u,w) \leq v \qtq{if, and only if,} w \leq f(u,v) \qquad\forall v\in B\can, \, w\in C\can. \]
In fact, this left adjoint can be restricted to the points of the dual Priestley spaces:
\begin{lem}\label{l:f-basics}
The following statements hold:
    \begin{enumerate}
        \item for all $w\in C\can$, $f\la_1(\ARG,w)$ is antitone and sends meets to joins;
        \item $f\la_1$ restricts to a map $\M(A\can)^\op \times \J(C\can) \to \J_0(B\can)$;
        \item for $x\in X$ and $z\in Z$, $f\la_1(x, \ki(z)) \neq 0$ iff $x \leq i(z)$.
    \end{enumerate}
\end{lem}
\begin{proof}
    Items 1 and 2 follow, respectively, by Proposition 4.1 and Theorem 4.4 in~\cite{GP2007}.
    
For item 3 note that, by Lemma~\ref{l:amp-f-g} and the definition of $i$, $f\la_1(x, \ki(z)) \nleq 0$ iff $\ki(z) \nleq f(x,0) = \zer\can x$ iff $\zer\can x \leq z$ iff $x\leq i(z)$.
\end{proof}

Recall that we assumed that Priestley duals are based on the sets of completely meet-irreducible elements of the respective canonical extensions. Therefore, Lemma~\ref{l:f-basics} shows that we can define the first partial operation
\[ - \colon Z\times X \rightharpoonup Y \]
with $\dom(-) = \{(z,x) \mid x \leq i(z)\}$, by setting\footnote{Observe that we decided to flip the arguments of $f\la_1$. This is because we typically think of the minus operation as being antitone in the second argument, whereas $f\la_1$ is antitone in the first argument.}
\[ z - x \ee{\coloneqq} \kappa(f\la_1(x, \ki(z))), \]
where $\kappa$ is the order-isomorphism between $\J(C\can)$ and $\M(C\can)$.

\begin{lem}\label{l:mip-topo-order-props}
    If $\dom(\mip)$ is seen as a subspace of $Z\times X^\op$, the following hold:
    \begin{enumerate}
        \item $\dom(\mip)$ is a closed up-set of $Z\times X^\op$;
        \item $\mip\colon \dom(\mip) \to Y$ is lower semicontinuous and monotone.
    \end{enumerate}
\end{lem}
\begin{proof}
    (1) \ $\dom(\mip)$ is an up-set because $i\colon Z\to X$ is monotone, and is closed as it coincides with the preimage of the closed subspace ${\geq}$ of $X\times X$ under the continuous map $i\times \mathrm{id}$.

    (2) \ Given $a\in A$, set $\w a \coloneqq \{x\in X \mid a\nleq x\}$. First, observe that $(-)\inv(\w a)$ is equal to the following open subset of $\dom(\mip)$:
    \[ \dom(\mip)\cap \bigcup_{b\in A} \w{h(b,a)} \times (X\setminus \w b). \]
    For the right-to-left inclusion, assume that $h(b,a)\nleq z$, $b\leq x$, and $x\leq i(z)$. Since $h(b,a) = f(b,a)$, we have $f(x,a)\nleq z$ iff $\ki z \leq f(x,a)$ iff $f\la_1(x,\ki z) \leq$ iff $a \nleq z - x$. Conversely, if $(z,x)\in \dom(\mip)$ then $a\leq x \mip x$ iff $f\la_1(x,\ki z)\leq a$ iff $f(x,a)\nleq z$. Since $x = \bigvee_i b_i$ for some non-empty subset $\{ b_i\}_i$ of $A$, and $f$ agrees with $g$ on open elements and $g$ distributes over non-empty joins, $f(x,a)\nleq z$ iff $\bigvee_i h(b_i,a) \nleq z$ iff $h(b_i,a) \nleq$ for some $i$. Consequently, $(z,x)$ belongs to $\w{h(b_i,a)} \times (X\setminus \w{b_i})$.

   This shows that $\mip\colon \dom(\mip) \to Y$ is lower semicontinuous. Monotonicity follows from item 1 of Lemma~\ref{l:f-basics} and the fact that $\kappa$ is monotone.
\end{proof}

We can recover $f$ just from the partial operation $-\colon \dom(-)\to Y$ as follows:
\begin{lem}\label{l:recovering-f}
    For every $u\in A\can$ and $v\in B\can$,
    \[ f(u,v) = \bigmee \{ z\in Z \mid \exists x\in X \text{ s.t. } u \leq x,\ (z,x)\in \dom(-) \text{ and }\, v \leq z - x \}. \]
\end{lem}
\begin{proof}
    For $(x,y,z)\in X\times Y\times Z$, we have
    \[
        f(x,y) \leq z
        \qtq{iff}
        \ki(z) \nleq f(x,y)
        \qtq{iff}
        f\la_1(x,\ki(z)) \nleq y.
    \]
   Moreover, $f\la_1(x,\ki(z)) \nleq y$ implies $x\leq i(z)$ because the latter inequation is equivalent to $f\la_1(x, \ki(z)) \neq 0$. We obtain that
    \[
        f(x,y) \leq z
        \qtq{iff}
        [\ki(z - x) \nleq y
        \ete{and}
        x \leq i(z)]
        \qtq{iff}
        [y \leq z - x
        \ete{and}
        x \leq i(z)].
    \]
Lastly, since $f(\ARG,\ARG)$ preserves arbitrary meets in both arguments, and every element of $A\can$ and $B\can$ is the meet of the completely meet-irreducible elements above it, $f(u,v) \leq z$ precisely when $x\leq i(z)$ and $y\leq z-x$ for some $(x,y)\in X\times Y$ satisfying $u \leq x$ and $v \leq y$.
\end{proof}

\subsection{The second partial operation}\label{s:duality-for-miss}
In the previous section, we presented a partial operation which mirrors the role of the pi extension $f$ of $h\colon A\times B\to C$ on the space side. Although any three-sorted partial operation $Z\times X \rightharpoonup Y$ satisfying the same topological and order-theoretic properties stated in Lemma~\ref{l:mip-topo-order-props} corresponds to a mapping $A\can \times B\can \to C\can$ that preserves meets in both coordinates, it does not necessarily follow that its restriction to $A\times B\to C$ is a double operator. This is the reason why we also need to compute the partial operation corresponding to the sigma extension $g\colon A\can \times B\can \to C\can$ of $h$.

This time, given an arbitrary $u\in A\can$, we do not have that $g(u,\ARG)\colon B\can \to C\can$ preserves~$0$. On the other hand, since $g$ is monotone, we know that $g(u,0)$ is the smallest element in the image of $g(u,\ARG)$, and so the restriction $g_1(u,\ARG)\colon B\can \to \upset g(u,0)$ preserves all joins and as such it has a right adjoint $g\ra_1$:
\[ g(u,v) \leq w \qtq{iff} g(u,0)\leq w \text{ and } v \leq g\ra_1(u, w) \qquad \forall v\in B\can, \,w\in C\can. \]

\begin{lem}\label{l:g-basics}
The following statements hold:
    \begin{enumerate}
        \item $g\ra_1$ is antitone in the first argument;
        \item  on its domain of definition, $g\ra_1$ is a partial map $\J(A\can)^\op \times \M(C\can) \rightharpoonup \M_1(B\can)$;
        \item for $x\in X$ and $z\in Z$, $g(\ki(x),0) \leq z$ iff $i(z)\nleq x$;
        \item for all $u\in A\can$ and $z\in Z$, $g\ra_1(u,z) \neq 1$.
    \end{enumerate}
\end{lem}
\begin{proof}
    Item~1 is easily verified, and item~2 follows from the proof of~\cite[Theorem 4.4]{GP2007}. 
    
For item 3, because $\ki(x)$ is a closed element, $g(\ki(x),0) = f(\ki(x), 0) = \zer\can (\ki(x))$ and so $\zer\can(\kappa \inv(x)) \leq z$ iff $\ki(x) \leq \zer\can{}\ra(z) = i(z)$ iff $i(z) \nleq x$, where the penultimate equivalence follows from Lemma~\ref{l:amp-f-g} and the definition of $i$. 
    
Finally, for item 4 we have $1 \leq g\ra_1(u,z)$ iff $1 = g(u,1) \leq z$. But this cannot be because $z$ is an element of $Z = \M(C\can)$, which does not contain 1.
\end{proof}

As before, we obtain a partial map
\[ \sim \colon Z\times X \rightharpoonup Y, \]
where $\dom(\sim) = \{(z,x) \mid i(z) \nleq x\}$, computed as follows:
\[ z \sim x \ee{\coloneqq} g\ra_1(\ki(x), z). \]

\begin{lem}
    If $\dom(\miss)$ is seen as a subspace of $Z\times X^\op$, the following hold:
    \begin{enumerate}
        \item $\dom(\miss)$ is an open up-set of $Z\times X^\op$;
        \item $\miss\colon \dom(\miss) \to Y$ is upper semicontinuous and monotone.
    \end{enumerate}
\end{lem}
\begin{proof}
    (1) \ By the same argument as in Lemma~\ref{l:mip-topo-order-props}, $\{ (z,x) \mid i(z)\leq x \}$ is closed and so its complement $\dom(\miss)$ is open. Further, $\dom(\miss)$ is an up-set because $i$ is monotone.

    (2) \ The argument is essentially the same as in Lemma~\ref{l:mip-topo-order-props}. Pick $a\in A$ and $(z,x)\in \dom(\miss)$, and let $\{b_i\}_i\sue A$ be such that $\ki x = \bigmee_i b_i$. Then, $a\leq x\miss z$ iff $g(\ki x, a)\leq z$ iff $\bigmee_i h(b_i, a) \leq z$ iff $h(b_i,a)\leq z$ for some $i$, where the second equivalence follows from the fact that $f$ and $g$ agree on closed elements of $A\can\times B\can$. Therefore,
    \[ (\miss)\inv(\w a) \ee= \dom(\miss)\cap \bigcap_{b\in B} \left((\w{h(b,a)}\times X) \cup (Z\times (X\setminus \w b))\right). \]
    The monotonicity of $\miss$ is a consequence of Lemma~\ref{l:g-basics} and the monotonicity of $\kappa$.
\end{proof}

Next, we show that $\sim$ is second-order-definable in terms of $-$. The proof follows the same ideas as the proof of~\cite[Proposition 3.14]{FGGM20}.

\begin{lem}\label{l:mis-from-mip}
    For all $(z,x)\in \dom(\sim)$,
    \[ z \sim x = \bigvee \{ z - x' \mid x' \in X \text{ s.t. } x' \nleq x \text{ and } (z,x') \in \dom(-)\}. \]
\end{lem}
\begin{proof}
    First, observe that for all $y\in \M(B\can)$,
    \begin{align}
        \ki(z \sim x) \leq \ki(y)
        & \ee\Longleftrightarrow \ki(y) \nleq z \sim x
        \notag
        \\
        & \ee\Longleftrightarrow g(\ki(x),\ki(y)) \nleq  z
        \notag
        \\
        & \ee\Longleftrightarrow f(\ki(x),\ki(y)) \nleq  z
        \notag
        \\
        & \ee\Longleftrightarrow \ki(z) \leq f(\ki(x),\ki(y))
        \notag
        \\
        & \ee\Longleftrightarrow f\la_1(\ki(x), \ki(z)) \leq \ki(y)
        \label{eq:sim-equiv-up}
    \end{align}
    where the third equivalence holds because $f$ and $g$ agree on open and closed elements of the product $A\can \times B\can$.

    Further, as every element $\ki(x)$ is equal to the meet of the elements of $X = \M(A\can)$ above it, and $f\la_1$ transforms meets into joins in the first argument (item 1 of Lemma~\ref{l:f-basics}),
    \begin{align*}
    f\la_1(\ki(x), \ki(z))
        &= \bigvee \{ f\la_1(x', \ki(z)) \mid x' \in X \text{ and } \ki(x)\leq x'\}
        \\
        &= \bigvee \{ \ki(z - x') \mid x'\in X,\ x' \nleq x \text{ and } (z,x')\in \dom(-) \}
    \end{align*}
    where the second equality holds because $f\la_1(x', \ki(z)) = 0$ whenever $(z,x') \notin \dom(-)$. Consequently, by equation~\eqref{eq:sim-equiv-up}, 
    \[
    \ki(z \sim x) \leq \ki(y) \ee \Longleftrightarrow \ee \ki(z - x') \leq \ki(y)
    \] 
    for every $x'\in X$ such that $x' \nleq x$ and $(z,x')\in \dom(-)$. The desired equality then follows because $\ki$ is an order-isomorphism.
\end{proof}

\section{\texorpdfstring{Extended Priestley duality: the case of $\Lh$}{Extended Priestley duality: the case of Gamma}}\label{s:apply-ext-prie-dual}
In this section, we exploit the theory exposed in Appendix~\ref{s:ext-prie-dual-can-ext} to provide a detailed justification of the definition of $\Lh$ and of its algebraic structure. As explained in Section~\ref{s:Lh}, we wish to obtain $\Lh$ as a codirected limit of finite chains $\Lh_n = \{0, \frac{1}{n}, \dots, 1\}$. Each such chain, with its natural order and the discrete topology, is a Priestley space. Its dual lattice is $(L_n)^\op$, where
\[ L_n \ee= \left(0 < \frac{1}{n} < \frac{2}{n} < \dots < 1 < \top\right). \]
Note that in Section~\ref{s:Lh}, we denoted by $L_n$ the order-dual of the lattice (of the same name) just defined. We make this change here to be able to view $\Lh_n$ as order-embedded into $L_n$ (\cf \ Remark~\ref{r:order-issues}). Observe that $L_n$ is isomorphic to its canonical extension because it is finite and hence complete. The image of the order-embedding $\Lh_n \mono L_n$ coincides precisely with the set $\M(L_n)$ of (completely) meet-irreducible elements of $L_n$.

Next, we define two binary operators, with the first $\oplus\colon L_n \times L_n \to L_n$ given by
\[ \top \oplus u = u \oplus \top = \top \quad \forall u\in L_n, \qtq{and} \frac{a}{n} \oplus \frac{b}{n} =
\begin{cases}
    \top & \text{ if } a + b > n \\
    \frac{a+b}{n} & \text{ otherwise}
\end{cases}
\]
and the second $\ominus\colon L_n \times L_n \to L_n$ by
\begin{align*}
    \top \ominus 0 &= \top
    &
    \top \ominus \frac{b}{n} &= \frac{n-b+1}{n} \quad\text{ (for $b\geq 1$)} \\
    \frac{a}{n} \ominus \frac{b}{n} &=
    \begin{cases}
        0 & \text{ if } b > a \\
        \frac{a-b}{n} & \text{ otherwise}
    \end{cases}
    &
    u \ominus \top &= 0 \quad \forall u\in L_n.
\end{align*}
In fact, these two operators are adjoint to each other: for all $u,v,w\in L_n$, $u \ominus v \leq w$ if, and only if, $u \leq v\oplus w$. Since $L_n$ is isomorphic to its canonical extension, $\oplus^\pi = \oplus^\sigma = \oplus$ and $\ominus^\pi = \ominus^\sigma = \ominus$.

\subsection{\texorpdfstring{Dualising $\oplus$ on $L_n$}{Dualising + on Ln}}
Observe that $\oplus$ is a three-sorted operator of type $L_n \times L_n \to L_n$. Because adjoints are uniquely determined, the left adjoint $f\la_1(u,\ARG)$ to $f(u,\ARG) = u \oplus (\ARG)$ is given by $f\la_1(u,v) = v\ominus u$. Then, by Lemma~\ref{l:f-basics}, the corresponding partial operation
\begin{align*}
    z - x = \kappa(f\la_1(x, \ki z)), \qtq{for} x \leq z,
    \label{eq:g_n-minus}
\end{align*}
is well defined. The order-isomorphism $\kappa\colon \J(L_n) \to \M(L_n)$ is given by $\top \mapsto 1$ and $\frac{a}{n} \mapsto \frac{a-1}{n}$, for $a > 0$. Consequently, it is not difficult to see that $\mip\colon \dom(\mip) \to \Lh_n$ is precisely the partial subtraction on $\Lh_n$.

\subsection{\texorpdfstring{Dualising $\ominus$ on $L_n$}{Dualising - on Ln}}
The assignment $f(u,v) = v \ominus u$ is a three-sorted operator of type $L_n \times L_n^\op \to L_n^\op$. Just observe that $u\ominus \top = 0 = 0 \ominus u$ and $\top \ominus 0 = \top$. Using Lemma~\ref{l:f-basics}, we see that the right adjoint $f\ra_1(u,\ARG)$ to $f(u,\ARG)$ restricts to $\M(L_n)^\op \times \J(L_n^\op) \to \J_0(L_n^\op)$, which is simply a monotone map $\M(L_n) \times \M(L_n) \to L_n$. Because right adjoints are unique, $f\ra_1(u,v) = u\oplus v$. Hence, the dual of $\ominus$ on $\Lh_n$ is the partial plus $x+z$, defined whenever $x \leq 1 - z$.

In this specific case, the formula in Lemma~\ref{l:mis-from-mip} to recover $\ominus$ from the partial plus looks as follows:
\begin{align*}
    f(u,v) = \bigvee \{ j\in \J(L_n)
        &\mid \exists x\in \M(L_n) \text{ s.t. } u\leq x,\ (x,\kappa(j)) \in \dom(+), \\
        &\quad\text{ and } \ki(x + \kappa(j)) \leq v \}.
\end{align*}

\subsection{Duals to floor and ceiling functions.}
Let $\epsilon\colon \Lh_n \mono \Lh_{nm}$ be the obvious inclusion. Floor and ceiling functions are right and left adjoint, respectively, to $\epsilon$. Just observe that, for all $\frac{a}{n}\in \Lh_n$ and $\frac{b}{nm}\in \Lh_{nm}$, we have
\[  \epsilon\left(\tfrac{a}{n}\right) \leq \tfrac{b}{nm} \ee\iff \tfrac{a}{n} \leq \lfloor \tfrac{b}{nm} \rfloor \qtq{and}  \lceil \tfrac{b}{nm} \rceil \leq \tfrac{a}{n} \ee\iff  \tfrac{b}{nm} \leq \epsilon\left(\tfrac{a}{n}\right). \]
Therefore, the floor function uniquely extends to the right adjoint of the lattice embedding $L_n \mono L_{nm}$. On the other hand, the ceiling function naturally extends to the left adjoint of the lattice embedding $\Lb_n \mono \Lb_{nm}$, with
\[ \Lb_n = \left(\bot < 0 < \frac{1}{n} < \dots < \frac{n-1}{n} < 1\right) \]
and $\Lb_{nm}$ defined similarly. Therefore, as a matter of convenience, we prefer to view ceiling functions as functions of type $\J(\Lb_{nm}) \to \J(\Lb_n)$ (\cf \ Section~\ref{s:priestley-duality}). To summarise, the duals of floor functions are the lattice embeddings
\[ i_{n,nm}\colon L_n \mono L_{nm}, \]
and the duals of ceiling functions are the lattice embeddings
\[ i'_{n,nm}\colon L_n \xrightarrow{~t~} \Lb_n \mono \Lb_{nm} \xrightarrow{~t\inv~} L_{nm}, \]
where $t\colon L_n \to \Lb_n$ is the renaming $\top \mapsto 1$, $0\mapsto \bot$ and $\frac{a}{n} \mapsto \frac{a-1}{n}$. The duals of floor and ceiling functions can also be computed in terms of plain Priestley duality, \cf \ Section~\ref{s:duality}.

Observe that $i_{n,m}\colon L_n \mono L_{nm}$ preserves $\oplus$ but not $\ominus$, because 
\[
i_{n,nm}(\top \ominus \tfrac{b}{n}) = i_{n,nm}(\tfrac{n-b+1}{n}) = \tfrac{n-b+1}{n}
\] 
whereas 
\[i_{n,nm}(\top) \ominus i_{n,nm}(\tfrac{b}{n}) = \tfrac{nm-b+1}{nm},
\] 
for $1\leq b \leq n$. A similar argument shows that $i'_{n,nm}$ does not preserve $\oplus$ nor $\ominus$.

The previous discussion justifies why we take the floor functions $\lfloor \ARG \rfloor\colon \Lh_{nm} \epi \Lh_n$ as the comparison maps in the codirected diagram that defines $\Lh$ and, moreover, why we equip the finite spaces $\Lh_n$ with partial minus, rather than partial plus.

\subsection{\texorpdfstring{The partial operations on $\Lh$}{The partial operations on Gamma}}\label{s:deriving-operations}
Saying that $\Lh$ is the codirected limit of the system of floor functions $\lfloor \ARG \rfloor\colon \Lh_{nm} \to \Lh_n$ is equivalent to saying that $\Lh$ is the Priestley space dual to the directed colimit of the lattice embeddings $L_n \mono L_{nm}$.
It is readily seen that this colimit can be identified with the lattice $\Lo = (\Q\cap\ui) \cup \{\top\}$, where $\top$ is the top element. Its canonical extension has the following properties:

\begin{lem}\label{l:Lo-can}
The following statements hold:
    \begin{enumerate}
        \item $\Lo\can$ can be identified with the lattice having underlying set
        \[ \{ 0\cc, 0\pp\} \cup \{ r\mm, r\pp \mid r\in (0,1)\setminus \Q\}\cup \{ q\mm, q\cc, q\pp \mid q\in (0,1)\cap \Q \} \cup \{1\mm, 1\cc, \top\}\]
       and the linear order defined by the following conditions:
 \begin{center}
\setlength{\tabcolsep}{1em}  
\begin{tabular}{ll}
(a) $r\mm \prec r\pp$ for all $r\in (0,1)\setminus \Q$ & (b) $q\mm \prec q\cc \prec q\pp$ for all $q\in (0,1)\cap \Q$  \\ 
 (c) $0\cc \prec 0\pp$ and $1\mm \prec 1\cc \prec \top$ & (d) $x\pp < y\mm$ if $x<y$ in $\ui$    
\end{tabular}
\end{center}
        where $\prec$ is the covering relation, \ie \ $\alpha \prec \beta$ if $\alpha < \beta$ and, whenever $\alpha \leq \gamma \leq \beta$, either $\alpha = \gamma$ or $\beta = \gamma$.

        \item $\M(\Lo\can) = \Lo\can \setminus (\{\top\}\cup \{ r\pp \mid r\in [0,1) \})$ and $\J(\Lo\can) = \Lo\can \setminus (\{0\cc\} \cup \{ r\mm \mid r\in (0,1] \})$.

        \item The order-isomorphism $\kappa\colon \J(\Lo\can) \to \M(\Lo\can)$ is given by
        \[ \top \mapsto 1\cc
           \qquad
           \qquad
           x\pp \mapsto
           \begin{cases}
               x\cc & \text{ if } x\in \Q \\
               x\mm & \text{ otherwise}
           \end{cases}
           \qquad
           \qquad
           q\cc \mapsto q\mm.
        \]
    \end{enumerate}
\end{lem}
\begin{proof}
 For item 1, it is easy to see that $\Lo\can$ is a complete lattice and that the embedding $\Lo \mono \Lo\can$, defined as $a \mapsto a\cc$ and $\top \mapsto \top$, satisfies the density and compactness conditions.
    Items 2 and 3 are easily verified and left to the reader.
\end{proof}

Because the lattice embeddings $L_n \mono L_{nm}$ preserve the operators $\oplus\colon L_n \times L_n\to L_n$, also $\Lo$ comes equipped with an operator $\oplus\colon \Lo \times \Lo \to \Lo$ obtained as the colimit of the corresponding operators on the finite sublattices $L_n$. This can be expressed as follows:
\[ \top \oplus u = u \oplus \top = \top, \quad \forall u\in \Lo, \quad \qtq{and}\quad \frac{a}{n} \oplus \frac{b}{m} =
\begin{cases}
    \top & \text{ if } am+ bn > n m \\
    \frac{am+bn}{nm} & \text{ otherwise.}
\end{cases}
\]
This time, though, $\oplus$ does not have a left adjoint because $\top \ominus \frac{a}{n}$ cannot be defined on~$\Lo$. This mirrors the fact that the inclusion maps $L_n \mono L_{mn}$ do not preserve $\ominus$ defined on finite lattices. Following the recipe of Section~\ref{s:duality-for-mip}, we compute the partial operation $-\colon \dom(-) \to \Lh$ from the left adjoint $a \leftplus (\ARG)$ of $a \oplus^\pi (\ARG)$. The pi extension $\oplus^\pi$ is commutative because so is $\oplus$. Therefore, the following lemma provides us with a full description of $\oplus^\pi \colon \Lo\can \to \Lo\can$.

\begin{lem}\label{l:table-opi}
Suppose $x,y\in \ui$. Then
\begin{equation*}
    x+y\leq 1
    \enspace
    \Rightarrow
    \hspace{-3em}
\begin{split}
&~ x\cc \opi y\cc \ee= (x+y)\cc
\\
    &\begin{rcases*}
        x\cc \opi y\mm \\
        x\mm \opi y\mm
    \end{rcases*}
    \ee= (x+y)\mm
    \end{split}
    \quad \text{and} \quad
    x+y< 1 \ \Rightarrow
    \begin{split}
    &\begin{rcases*}
        x\pp \opi y\pp \\
        x\pp \opi y\cc \\
        x\pp \opi y\mm
    \end{rcases*}
    \ee= (x+y)\pp,
    \end{split}
\end{equation*}
whenever the expressions make sense. In all the remaining cases, we get $\top$.
\end{lem}
\begin{proof}
Suppose $x+y\leq 1$. It is not difficult to see that the set of open elements of $\Lo\can$ is $\I(\Lo\can)=\M_1(\Lo\can)=\M(\Lo\can)\cup\{\top\}$. Just observe that every proper ideal in a linearly ordered lattice is prime. Hence, by item 2 in Lemma~\ref{l:Lo-can}, 
\[
\I(\Lo\can)=\{ 0\cc\} \cup \{ r\mm \mid r\in (0,1)\setminus \Q\}\cup \{ q\mm, q\cc \mid q\in (0,1)\cap \Q \} \cup \{1\mm, 1\cc, \top\}.
\]
By the definition of pi extensions on open elements (\cf \ Section~\ref{s:priestley-duality}), we get
\[
x\cc \opi y\cc = \bigvee{\{a\oplus b\mid a,b\in \Lo, \ a\leq x\cc \text{ and } b\leq y\cc\}}=(x+y)\cc.
\]
The same equation also tells us the results of $x\cc \opi y\mm$ and $x\mm \opi y\mm$, since all the involved elements are open:
\begin{align*}
x\cc \opi y\mm
    &=\bigvee{\{a\oplus b\mid a,b\in \Lo, \ a\leq x\cc \text{ and } b\leq y\mm\}}
    \\
    &=\bigvee{\{x\oplus b\mid b\in \Q\cap\ui, \ b< y\}}=(x+y)\mm,
\end{align*}
and
\[
x\mm \opi y\mm=\bigvee{\{a\oplus b\mid a,b\in \Q\cap\ui, \ a< x \text{ and } b< y\}}=(x+y)\mm.
\]
Suppose now that $x+y<1$. The definition of pi extensions on possibly non-open elements, along with the cases covered in the first part of the proof, yield 
\begin{align*}
x\pp\opi y\pp &=\bigwedge{\{w\opi z\mid w,z\in  \I(\Lo\can), \ x\pp\leq w \text{ and } y\pp\leq z\}} \\
&=\bigwedge{\{w\mm\opi z\mm\mid w,z\in (0,1], \ x< w \text{ and } y< z\}} \\
&=\bigwedge{\{(w+ z)\mm\mid w,z\in (0,1], \ x< w \text{ and } y< z\}}=(x+y)\pp.
\end{align*}
All the remaining cases are computed similarly.
\end{proof}

Since the assignment $a \mapsto a \oplus 0$ is equal to the identity, its dual $i\colon \Lh \to \Lh$ is also the identity. Therefore, $\dom(\mip)$ is equal to $\{ (z,x) \mid x\leq z \}$. Moreover, it follows from Lemma~\ref{l:Lo-can}, along with Lemma~\ref{l:table-ompi} below, that the partial operation $z \mip x = \kappa(x \leftplus \ki z)$ is exactly the one described in Section~\ref{subs:mip-and-miss}.

\begin{lem}\label{l:table-ompi}
Let $u,v\in \Lo\can$. If $u\leq v$, then $u\leftplus v=0\cc$. Otherwise, for $x,y\in \ui$,
\[
\begin{array}{|c|c|c|c|}
\hline
\leftplus & y\cc & y\mm & y\pp \\
\hline
x\cc & (x-y)\cc & (x-y)\pp & (x-y)\mm \\
\hline
x\mm & (x-y)\mm & (x-y)\mm & (x-y)\mm \\
\hline
x\pp & (x-y)\pp & (x-y)\pp & (x-y)\mm \\
\hline
\end{array}
\]
whenever the expressions make sense (read as: $(\text{row})\leftplus(\text{colon})$). Finally,
\[
\top\leftplus x\cc=\top\leftplus x\mm=(1-x)\pp \ \text{ and } \ \top\leftplus x\pp=(1-x)\mm.
\]
\end{lem}
\begin{proof}
    Follows by a routine computation using the usual description of left adjoints in terms of infima, \ie \ $u\leftplus v=\bigwedge{\{w\in \Lo\can\mid u\leq v\opi w\}}$ for every $u,v\in\Lo\can$.
\end{proof}

Lastly, we compute the expression for the second partial operation $\miss\colon \dom(\miss)\to \Lh$. However, instead of computing the sigma extension of $\oplus$ and its (partial) right adjoint, as in Section~\ref{s:duality-for-miss}, we use the formula from Lemma~\ref{l:mis-from-mip}. Because $i\colon \Lh \to \Lh$ is the identity mapping, $\mbox{\dom(\miss)} = \{ (z,x) \mid z \nleq x \} = \{ (z,x) \mid x < z \}$. Furthermore, by Lemma~\ref{l:mis-from-mip}, for every $(z,x)\in \dom(\sim)$ we have 
\[ z \sim x = \bigvee \{ z - x' \mid x' \in \Lh \text{ s.t. } x < x' \leq z \}. \]

Now, since $x < z$, there are two possible cases. Either $x = q\mm < q\cc = z$ for some rational $q\in (0,1]$, in which case the unique $x'$ such that $x < x' \leq z$ is $q\cc$, or $x$ is not an immediate predecessor of $z$. In the latter case, by a similar argument, for every $x < x' \leq z$ there is a rational $q\in (0,1]$ such that $x < q\cc \leq x'$. Thus, because $\mip$ is antitone in the second coordinate, we obtain exactly the same expression as used in Section~\ref{subs:mip-and-miss} to define $\miss$:
\[ z \sim x = \bigvee \{ z - x' \mid q \in (0,1]\cap\Q \text{ s.t. } x < q\cc \leq z \}. \]
The only difference is that $z \miss x$, as described in Section~\ref{subs:mip-and-miss}, is also defined for $z = x$.

\bigskip
To conclude, we offer a proof of Lemma~\ref{lem:mip-is-right-adjoint}, which relies on the theory exposed above.
\begin{proof}[Proof of Lemma~\ref{lem:mip-is-right-adjoint}]
    Let $f$ be the pi extension $\oplus^\pi$ of $\oplus$. Then,
    \[ f(x,y) \leq z \iff \ki z \nleq f(x,y) \iff f\la_1(x,\ki z) \nleq y \iff y \leq z \mip x \]
    for every $y\in \Lh$ and $(z,x)\in \dom(\mip)$. By Lemma~\ref{l:table-opi}, the restriction of $f$ to
    \[ (x\cc,y\cc), (x\cc, y\mm), (x\mm, y\cc),\text{ and } (x\mm,y\mm) \]
    where $x+y\leq 1$ (whenever the expressions make sense), is a well defined partial operation on $\Lh$. Observe that this domain of definition is precisely $\{ (x,y) \mid x \leq 1\cc \mip y\}$ because $x \leq 1\cc \mip y$ iff $f(x,y) \leq 1\cc$ iff $f(x,y) \neq \top$.
\end{proof}

\end{document}